\documentclass[%
 aps,floatfix
 amsmath,amssymb,
 reprint,%
]{revtex4-2}
\usepackage[utf8]{inputenc}
\usepackage[english]{babel}
\usepackage[T1]{fontenc}
\usepackage{amsmath}
\usepackage{hyperref}
\usepackage{mathptmx}
\usepackage{tikz}
\usepackage{lipsum}
\usepackage{graphicx} 
\usepackage{dcolumn} 
\usepackage{bm} 
\usepackage{xcolor}
\usepackage{colortbl}
\usepackage{latexsym}
\usepackage{caption}
\usepackage{subcaption}
\usepackage{parskip}
\usepackage{braket}
\usepackage{physics}
\usepackage{cancel}
\usepackage{balance}
\usepackage{amsthm}
\usepackage{multirow}
\usepackage{etoolbox}
\usepackage{dsfont}
\usepackage{nicematrix}
\usepackage{qcircuit}

\allowdisplaybreaks

\usepackage{algorithm}

\usepackage{algpseudocode}

\begin{document}

\title{Efficient application of the factorized form of the unitary coupled-cluster ansatz for the variational quantum eigensolver algorithm by using linear combination of unitaries 
}

\author{Luogen Xu}%
 \email{lx63@georgetown.edu}
\affiliation{Department of Physics, Georgetown University, 37$^{\rm th}$ and O Sts. NW, Washington, DC 20057 USA}

\author{J. K. Freericks}%
\affiliation{Department of Physics, Georgetown University, 37$^{\rm th}$ and O Sts. NW, Washington, DC 20057 USA}

\date{\today}

\begin{abstract}
The variational quantum eigensolver  is one of the most promising algorithms for near-term quantum computers. It has the potential to solve quantum chemistry problems involving strongly correlated electrons, which are otherwise difficult to solve on classical computers. The variational eigenstate is constructed from a number of factorized unitary coupled-cluster terms applied onto an initial (single-reference) state. Current algorithms for applying one of these operators to a quantum state require a number of operations that scales exponentially with the rank of the operator. We  exploit a hidden SU($2$) symmetry to allow us to employ the linear combination of unitaries approach, Our  \textsc{Prepare} subroutine uses $n+2$ ancilla qubits for a rank-$n$ operator. Our \textsc{Select}($\hat U$) scheme uses $\mathcal{O}(n)$ \textsc{Cnot} gates. This results in an full algorithm that scales like the cube of the rank of the operator $n^3$, a significant reduction in complexity for rank five or higher operators. This approach, when combined with other algorithms for lower-rank operators (when compared to the standard implementation, will make the factorized form of the unitary coupled-cluster approach much more efficient to implement on all types of quantum computers.
\end{abstract}

\maketitle

\section{Introduction}
One of the important motivations for developing quantum computers is their potential to simulate strongly correlated many-body systems efficiently \cite{aspuru-guzik_2005, whitfield_biamonte_aspuru-guzik_2011}. Algorithms that exactly diagonalize the electronic Hamiltonian, known as the full configuration interaction approach, scale exponentially with the size of the Hilbert space, making it applicable to very few cases \cite{fci} on classical computers. The configuration interaction (CI) method offers an approximate solution by truncating the Hilbert space to only include the most important basis states. However, the energy calculated by the CI method does not scale properly with the size of the system when used on molecules with varying sizes, nor does it predict the dissociation energy correctly because it cannot produce factorized atomic states. The coupled cluster (CC) method addresses these issues by being both size consistent and size extensive. It is also memory efficient because it does not explicitly construct the energy eigenstate. Instead, the set of amplitudes for the CC ansatz is calculated iteratively by the so-called amplitude equations \cite{bartlett_purvis_1978, cc, purvis_bartlett_1982}, which correspond to zeroing out the row (or column) of the Hamiltonian matrix that corresponds to the initial single-reference state. The CC method with single, double and (perturbative) triple excitations is regarded as the ``gold standard'' for computational chemistry \cite{aspuruguzik_uccreview}.

Quantum computers have been proposed as being capable of solving a set of quantum chemistry problems that are otherwise difficult or very challenging on classical machines: namely, molecules that contain both weakly and strongly correlated electrons. One of the most promising algorithms for the noisy intermediate-scale quantum (NISQ) era is the variational quantum eigensolver (VQE), where the trial wave function is prepared on the quantum hardware and the expectation value of the energy is measured there as well; the parameters in the eigenstate are optimized variationally on classical machines \cite{vqe, preskill_2018}. The conventional coupled-cluster ansatz is given as $\ket{\psi_{CC}}=e^{\hat T}\ket{\psi_{ref}}$, where $\ket{\psi_{ref}}$ is a trial wave function (often chosen to be the single-reference Hartree-Fock state), and $\hat T=\sum_{k=1}^n \hat T_k$ is the cluster operator consisting of up to rank-$n$ excitations ($n$ electrons are removed from the Hartree-Fock state and replaced by $n$ electrons in virtual orbitals). The excitation operator is given as 
\begin{equation}
\hat T_k =\frac{1}{(k!)^2}\sum_{ij\cdots} ^{occ} \sum_{ab\cdots} ^{vir} t_{ij\cdots} ^{ab\cdots} \hat A_{ij\cdots} ^{ab\cdots}, 
\end{equation}
and $\hat A_{ij\cdots} ^{ab\cdots} = \hat a_a^\dagger \hat a_b^\dagger \cdots \hat a_j \hat a_i$, where $\hat a_a^\dagger$ is the creation operator acting on virtual orbital $a$ and $\hat a_i$ is the annihilation operator acting on occupied orbital $i$. Traditionally, the CC method employs a similarity-transformed Hamiltonian to obtain a set of equations to determine the amplitudes $t$:
\begin{align}
    \bra{\psi_{ref}} e^{-\hat T}\hat H e^{\hat T} \ket{\psi_{ref}} = E \label{eq:amplitude1}\\
    \bra{\psi_\mu} e^{-\hat T}\hat H e^{\hat T} \ket{\psi_{ref}} = 0 \label{eq:amplitude2}
\end{align}
where $\bra{\psi_\mu} = \bra{\psi_{ref}}\hat A_\mu$. In practice, this set of amplitude equations is solved iteratively, which yields the energy without needing to construct the energy eigenstate. The total number of amplitude equations is given by the number of amplitudes in the expansion of the $\hat T$ operator, which is much smaller than the total number of Slater determinants in the $\ket{\psi_{CC}}$ (which is typically exponentially larger). The properties of size consistency and size extensivity for the CC ansatz stem from the facts that the similarity-transformed Hamiltonian $e^{-\hat T}\hat H e^{\hat T}$ is additively separable and the term $e^{T}$ is multiplicatively separable. Notice that the electronic Hamiltonian for the molecule (in second quantization) is given by
\begin{equation}
    H = \sum_{ij} h_{ij}\hat a_i^\dagger \hat a_j +\frac{1}{2}\sum_{ijkl}g_{ijkl} \hat a_i^\dagger\hat a_j^\dagger \hat a_k \hat a_l,
\end{equation}
where $h_{ij}$ are the one-electron integrals and $g_{ijkl}$ are the two-electron integrals:
\begin{align}
    h_{ij} = \int dr_1 \phi_i^* (r_1) \Bigg(-\frac{1}{2}\nabla_{r_1} ^2 - \sum_{I=1} ^M \frac{Z_I}{R_{1I}} \Bigg)\phi_j (r_1) \\
    g_{ijkl} = \int dr_1 dr_2 \phi_i^*(r_1)\phi_j^*(r_2)\frac{1}{r_{12}}\phi_k(r_1)\phi_l(r_2).
\end{align}
Here, $M$ is the number of atoms in the system, $Z_I$ are their atomic numbers, $R_{1I} = \abs{r_1 - R_I}$, $r_{12} = \abs{r_1 - r_2}$, and $\phi(r)$ are the single-particle optimized orbitals from the   HF solution \cite{szabo_ostlund_2006, taketa_huzinaga_o-ohata_1966}. In order to solve the amplitude equations (\ref{eq:amplitude1}) and (\ref{eq:amplitude2}), we need to explicitly compute the similarity-transformed Hamiltonian. Using the Hadamard lemma, we can rewrite the transformed Hamiltonian as
\begin{align}
    e^{-\hat T}\hat H e^{\hat T} = &\hat H + [\hat H, \hat T] + \frac{1}{2!} [[\hat H, \hat T],\hat T] + \frac{1}{3!}[[[\hat H, \hat T], \hat T],\hat T], \hat T] \nonumber\\
    + &\frac{1}{4!}[[[[\hat H, \hat T], \hat T], \hat T], \hat T] + \cdots \label{eq:bch}
\end{align}
Conveniently, the series truncates at the fourth order due to the Hamiltonian having only one- and two-body interaction terms \cite{shavitt_bartlett_2009, cc} and the excitations always being from real to virtual orbitals. Traditionally, this projective method to determine the CC amplitudes is preferred over variational methods due to the non-unitarity of the $e^{\hat T}$ operator \cite{aspuruguzik_uccreview, cc}. 

Despite its success, the lack of unitarity prevents the CC operators to be implemented on quantum computers. This suggests using the unitary coupled-cluster ansatz (UCC), whose cluster operator now includes the excitation \textit{minus} the deexcitation operator $\hat T - \hat T^\dagger$ \cite{bartlett_kucharski_noga_1989, schaefer_2013}. Similar to the CC approximation, only the low-rank cluster operators such as singles and doubles are usually selected for the variational eigenstate ansatz; but for more strongly correlated systems, one expects that higher-rank factors will also be needed. In practice, a projective method like the one used in the CC calculation does not work with the UCC ansatz because the similarity-transformed Hamiltonian no longer truncates after the fourth term. Common strategies for carrying it out on classical computers include truncating the Hadamard lemma series at a fixed order \cite{bartlett_kucharski_noga_1989}, expanding the exponential operator in a power series and then truncating it when the higher-rank terms no longer change the eigenfunction \cite{cooper_knowles_2010}, and using an exact operator identity of the factorized form of the UCC to allow the wavefunction to be constructed in a tree structure \cite{chen_cheng_freericks_2021}. But, there exists no simple method to work directly with the UCC ansatz in its original form. Since we are working with non-commuting fermionic operators $\hat a_a^\dagger \hat a_b^\dagger \cdots \hat a_j \hat a_i - \hat a_i^\dagger \hat a_j^\dagger \cdots \hat a_b \hat a_a$ in the exponent, one common way to decompose such a function is to adopt a Trotter product formula:
\begin{equation}
    \hat U_{UCC} = e^{\sum_k \theta_k (\hat T_k - \hat T_k^\dagger)} = \lim_{N\rightarrow \infty}\Bigg(\prod_k e^{\frac{\theta_k}{N}(\hat T_k - \hat T_k^\dagger)} \Bigg)^N.
    \label{eq:ucctrotter}
\end{equation}
Another useful method is to express the ansatz in a factorized form, given by
\begin{equation}
    \hat U^\prime_{UCC} = \prod_k e^{\theta_k (\hat T_k - \hat T_k^\dagger)},
\end{equation}
which corresponds to the first-order approximation of the Trotter product formula in Eq.~(\ref{eq:ucctrotter}). The benefit of only using the $N=1$ extreme case is two-fold: the quantum resources required to prepare the factorized UCC ansatz are much smaller than higher-order approximations and the Trotter errors of the first-order approximation can be ameliorated by the fact that the calculation is variational \cite{romero_babbush_mcclean_hempel_love_aspuru-guzik_2018, barkoutsos_2018}. Within the classical computational chemistry framework, work by Chen, \textit{et al.} \cite{chen_cheng_freericks_2021} created an algorithm using the factorized form of the UCC that produces significantly better results for strongly correlated systems and comparable results in terms of accuracy for weakly correlated systems.

To implement the factorized UCC ansatz on quantum computers, one needs to transform the cluster operators $\hat T - \hat T^\dagger$ expressed in the fermionic language into a spin language (via the Jordan-Wigner transformation, or other fermionic encodings). A common realization of this approach is to exactly simulate the  individual exponentials of Pauli strings found after the JW transformation of $e^{\hat T - \hat T^\dagger}$ \cite{romero_babbush_mcclean_hempel_love_aspuru-guzik_2018, barkoutsos_2018}. This is possible because the different $2^{2n-2}$ Pauli strings (for a rank-$n$ UCC factor) commute with each other. In our previous work, we found a way of reducing the number of control-NOT (\textsc{Cnot}) gates in quantum circuits for the factorized UCC ansatz by introducing extra ancilla qubits \cite{xu_lee_freericks_2022}, with the largest reductions for the higher-rank factors. In this work, we introduce a method to directly simulate the sum of terms obtained from a hidden SU($2$) symmetry of the first-order Trotter product that greatly reduces the number of multi-qubit entanglement gates of factorized UCC circuits.

\section{Background}

\subsection{SU(2) identity for individual UCC factors}

Recall the rank-$n$ cluster operator is defined as
\begin{equation}
    \hat T_k =\frac{1}{(k!)^2}\sum_{ij\cdots} ^{occ} \sum_{ab\cdots} ^{vir} \theta_{ij\cdots} ^{ab\cdots} \Big(\hat A_{ij\cdots} ^{ab\cdots}-\hat A_{ab\cdots} ^{ij\cdots} \Big).
\end{equation}
The first two ranks are
\begin{align}
    \hat T_1 &= \sum_{ia}\theta_i ^a \big( \hat a_a ^\dagger \hat a_i - \hat a_i ^\dagger \hat a_a\big)  = \sum_{ia} \theta_i^a \big(\hat A_i^a - \hat A_a^i\big)\\
    \hat T_2 &= \frac{1}{2}\sum_{ijab} \theta_{ij}^{ab} \big(\hat a_a ^\dagger \hat a_b ^\dagger \hat a_j \hat a_i-\hat a_i ^\dagger \hat a_j ^\dagger \hat a_b \hat a_a\big) \nonumber\\
    &= \frac{1}{2}\sum_{ijab} \theta_{ij}^{ab} \Big(\hat A_{ij}^{ab} - \hat A_{ab}^{ij}\Big),
\end{align}
where $\hat a_a^{\dagger}$ is the fermionic creation operator on the virtual orbital $a$ and $\hat a_i$ is the fermionic annihilation operator on the real orbital $i$, and they obey the standard anti-commutation relations given by
\begin{equation}\label{ccr}
    \{\hat a_i, \hat a_j\} = 0; \{\hat a_i^{\dagger}, \hat a_j^{\dagger}\} = 0; \{\hat a_i, \hat a_j^\dagger\} = \delta_{ij} 
\end{equation}
where $\{A, B\}=AB+BA$ and $\delta_{ij}$ is the Kronecker delta function. First, we note that because $\{i,j,k,\cdots\}$ and $\{a,b,c,\cdots\}$ are disjoint sets, $\hat{A}^2=0=\hat{A}^{\dagger 2}$, so the squared term becomes
\begin{align}
    (\hat A - \hat A^\dagger)^2 &= -\hat{A}\hat A^\dagger - \hat A^\dagger \hat{A} \nonumber \\
    &= -\hat n_{a_1}\hat n_{a_2}\cdots\hat n_{a_n}(1-\hat n_{i_1})(1-\hat n_{i_2})\cdots(1-\hat n_{i_n}) \nonumber\\ 
    & -(1-\hat n_{a_1})(1-\hat n_{a_2})\cdots(1-\hat n_{a_n})\hat n_{i_1}\hat n_{i_2}\cdots\hat n_{i_n}, 
    \label{eq:square_identity}
\end{align}
where $\hat n_\alpha = \hat a_\alpha ^\dagger \hat a_\alpha$ is the number operator for spin-orbital $\alpha$. The cubed term then becomes
\begin{equation}
    (\hat A-\hat A^\dagger)^3 =\hat A\hat A^\dagger\hat A-\hat A^\dagger\hat A\hat A^\dagger=\hat A-\hat A^\dagger,
    \label{eq:cube_identity}
\end{equation}
because the projection operators $\hat{n}$ and $1-\hat{n}$ evaluate to one when they act on the corresponding fermionic operators.
For any UCC factor, the power series expansion is given as
\begin{equation}
    e^{\theta (\hat A - \hat A^\dagger)} = \sum_{n=0}^{\infty} \frac{\theta ^n}{n!} (\hat A - \hat A^\dagger)^n.
    \label{eq:uccpowerseries}
\end{equation}
Combining with equations (\ref{eq:square_identity}) and (\ref{eq:cube_identity}), we can then exactly write the sum as
\begin{align}
    e^{\theta (\hat A - \hat A^\dagger)} = &\hat I + \sin{\theta}(\hat A - \hat A^\dagger)
    +(\cos{\theta} - 1)(\hat n_{a_1}\hat n_{a_2}\cdots\hat n_{a_n}\nonumber\\
    &\times(1-\hat n_{i_1})(1-\hat n_{i_2})\cdots(1-\hat n_{i_n}) +(1-\hat n_{a_1})\nonumber\\
    &\times(1-\hat n_{a_2})\cdots(1-\hat n_{a_n})\hat n_{i_1}\hat n_{i_2}\cdots\hat n_{i_n}),
    \label{eq:uccidentity}
\end{align}
for any given set of occupied orbitals $\{i_1\cdots i_n \}$ and virtual orbitals $\{a_1\cdots a_n \}$ of rank $n$ \cite{xu_lee_freericks_2020, freericks_2022, chen_cheng_freericks_2021}. This identity gives a clear picture of what is happening after a UCC factor is applied to a state. If the state is neither excited by $\hat A$ nor deexcited by $\hat A^\dagger$, the state is unchanged by the UCC factor. Otherwise, the UCC factor acting on the state is equivalent to a cosine multiplied by the original state plus a sine multiplied by the excited (or deexcited) state, just as we would expect from a rotation in the many-body configuration space. 
\subsection{Jordan-Wigner transformation of the SU(2) identity}
Hamiltonians written in fermionic terms need to be re-expressed in terms of spin operators in order to be implemented by quantum computers. In this work, we choose to work with the JW transformation for the fermionic encoding. This transformation is given by
\begin{align}
    \hat a_k &= \frac{1}{2}(X_k + iY_k)\otimes Z_{k+1}\otimes Z_{k+2}\otimes\cdots\otimes Z_N\label{eq:jw1}\\
    \hat a_k ^\dagger &= \frac{1}{2}(X_k - iY_k)\otimes Z_{k+1}\otimes Z_{k+2}\otimes\cdots\otimes Z_N\label{eq:jw2}\\
    \hat n_k &= \hat a_k ^\dagger \hat a_k = \frac{1}{2}(1-Z_k)\label{eq:jw3},
\end{align}
where $ X$, $ Y$, and $ Z$ are the standard Pauli matrices, and $0 \leq k \leq N-1$, for the $N$ qubits that describe the molecule. The qubit state $|0\rangle$ has no electrons and $|1\rangle$ has one electron. The SU($2$) identity for a UCC factor, as shown in Eq.~ (\ref{eq:uccidentity}), can be reexpressed in terms of the Pauli operators using Eqs.~(\ref{eq:jw1}), (\ref{eq:jw2}), and (\ref{eq:jw3}). For a factorized UCC double (UCCD) operator, the transformation is as follows
\begin{align}
    \hat U(\theta) = &\exp\Big(\theta \big(\hat a_i ^\dagger \hat a_j ^\dagger \hat a_k \hat a_l - \hat a_l ^\dagger \hat a_k ^\dagger \hat a_j \hat a_i\big) \Big) \nonumber\\
    = &\hat I + \sin{\theta}(\hat a_i ^\dagger \hat a_j ^\dagger \hat a_k \hat a_l - \hat a_l ^\dagger \hat a_k ^\dagger \hat a_j \hat a_i) + (\cos{\theta}-1)\nonumber\\
    &\times\big(\hat n_l \hat n_k (1-\hat n_i)(1-\hat n_j) + (1-\hat n_l)(1-\hat n_k)\hat n_i \hat n_j\big)\nonumber\\
    = & \hat I + \frac{i\sin{\theta}}{8} \bigotimes _{a=l+1} ^{k-1} Z_{a}\bigotimes _{b=j+1} ^{i-1} Z_{b} \times \nonumber\\
    \bigg( &X_l  X_k  Y_j  X_i +  Y_l  X_k  Y_j  Y_i + X_l  Y_k  Y_j  Y_i +  X_l  X_k  X_j  Y_i \nonumber \\
    - &Y_l  X_k  X_j  X_i -  X_l  Y_k  X_j  X_i - Y_l  Y_k  Y_j  X_i -  Y_l  Y_k  X_j  Y_i\bigg) \nonumber\\
    + &\frac{1}{8}(\cos{\theta}-1)(\hat I + Z_i Z_j + Z_l Z_k - Z_j Z_l - Z_j Z_k\nonumber\\
    - &Z_i Z_l - Z_i Z_k + Z_i Z_j Z_k Z_l).
    \label{eq:su2jw}
\end{align}
Note that the JW strings simplify, because $Z_k^2=\mathbb{I}$ for all cases where two strings overlap. This expression is a unitary operator, but it is also here expressed as a sum over unitary operators, because Pauli strings are both Hermitian and unitary.

\subsection{Linear combination of unitaries}
To simulate the sum in Eq.~(\ref{eq:su2jw}) on a quantum computer, we use the linear combination of unitaries (LCU) query model  \cite{childs_wiebe_2012, berry_childs_cleve_kothari_somma_2015}. Given an input operator $\hat{U}$ represented by a sum of unitaries $\hat U = \sum_n \alpha_n \hat U_n$, with coefficients $\alpha_n$ for each unitary operator $\hat U_n$, the LCU technique will create a circuit to evaluate this operator acting on a state. It first prepares an ancilla bank with coefficients based on the coefficients in the linear combination:
\begin{equation}
    \hat{B}\ket{0} = \frac{1}{\sqrt{s}} \sum_n \sqrt{\alpha_n} \ket{n}.
    \label{eq:lcuprepare}
\end{equation}
Here, $\frac{1}{\sqrt{s}}$ is a normalization factor, $\ket{0}$ is the initial state of the ancilla bank, and $\ket{n}$ is the product state that will later encode the unitaries in the LCU procedure. The operator \textsc{Select}($\hat U$) is then used to create entanglement between the ancilla bank and system states
\begin{equation}
    \textsc{Select}(U)\ket{n}\otimes\ket{\psi} = \ket{n}\otimes U_n\ket{\psi}.
    \label{eq:selectdef}
\end{equation}
One of the hallmarks of the LCU approach is that if the original operator $\hat U$ is unitary and $s\leq 2$, then a single step of oblivious amplitude amplification is able to exactly apply the $\hat U$ to the state \cite{childs_wiebe_2012}. Note that in our case the UCC factor, given in Eq.~(\ref{eq:su2jw}) is unitary and $s = \cos{\theta} + \sin{\theta} \leq 2$ for all $\theta$, so it always satisfies this criteria. Hence, the LCU treatment of the sum is exact. The oblivious amplitude amplification is given by
\begin{equation}
    -\hat{W}\hat{R}\hat{W}^\dagger \hat{R}\hat{W}\ket{0}\otimes\ket{\psi} = \ket{0}\otimes\hat{U}\ket{\psi},
\end{equation}
where the $\hat{W}$ and $\hat{R}$ operators are defined as
\begin{align}
    \hat{W} := &(\hat{B}^\dagger\otimes 1) \textsc{Select}(\hat U)(\hat{B}\otimes 1), \nonumber\\
    \hat{R} := &1 - 2(\ket{0}\bra{0}\otimes 1).
\end{align}
The main source of circuit complexity of the LCU query model comes from the unitary transformation $\hat{W}$ because it involves applying \textsc{Select}($\hat U$), which itself can contain a substantial number of multi-qubit gates and quickly outgrows the capability of near-term hardware. One efficient circuit implementation of the \textsc{Select}($\hat U$) subroutine for a generic fermionic Hamiltonian uses $\mathcal{O}(\eta)$ Clifford and $T$ gates, with Clifford gates running in $\mathcal{O}(\log^2 \eta)$ layers and $T$ gates in $\mathcal{O}(\eta)$ layers.  Here, $\eta$ is the number of spin orbitals \cite{wan_2021}. The ancilla preparation operator $\hat{B}$ is often implemented by rotations and controlled rotations on the target qubits, followed by $\text{Hadamard} ^{\otimes \eta}$ gates that create the required entanglement state for the ancilla bank.

\section{Circuit construction}\label{sec:circuitexample}
We begin by illustrating the circuit implementation of the \textsc{Prepare} and \textsc{Select}($\hat U$) subroutines present in the LCU adaptation of the UCC factors for doubles. The doubles are the most ubiquitous terms in the low-rank representation of a UCC ansatz. Later in this section, we will show that UCC factors of arbitrary rank-$n$ can be implemented via a similar algorithm. High-rank factors are necessary to generate an accurate correlation energy in strongly correlated systems.
\subsection{\textsc{Prepare} subroutine}\label{sec:prepare}
\begin{figure}[ht!]
    \centering
    \includegraphics[width=0.45\textwidth]{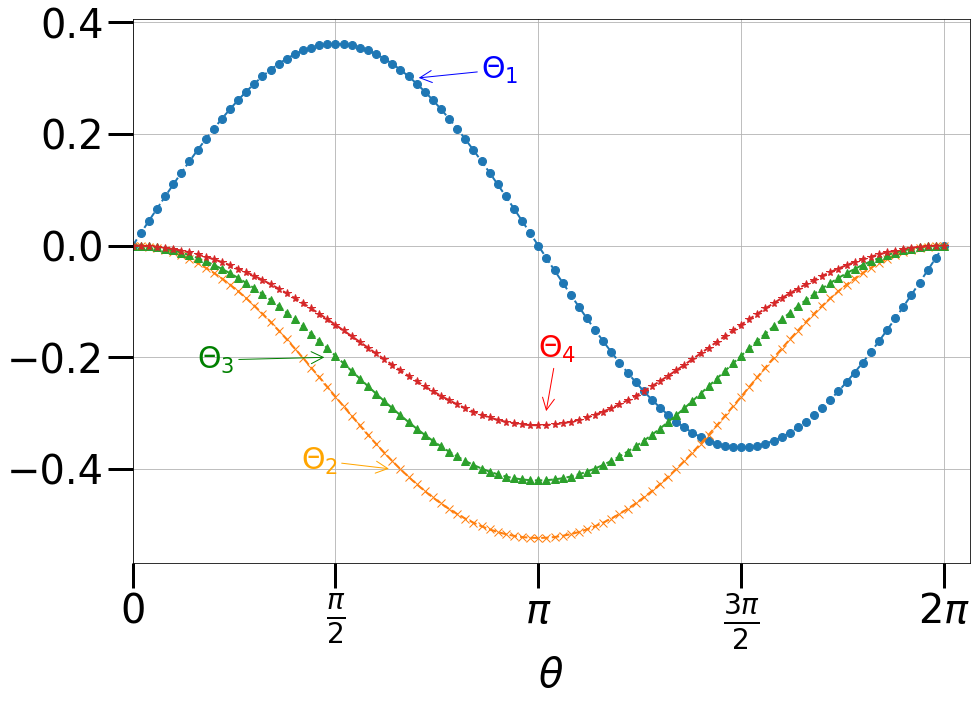}
    \caption{The four angles used in the \textsc{Prepare} subroutine for a UCC doubles factor with an amplitude given by $\theta$.}
    \label{fig:uccd_prepare_angles}
\end{figure}

The unitary transformation $\hat{B}$ is used to generate required entangled state in the ancilla bank, shown in Eq.~(\ref{eq:lcuprepare}). The operator in Eq.~(\ref{eq:su2jw}), lends itself to a binary encoding, where we create the linear combination of states multiplied by amplitudes: that is, the sum of $\alpha_1\ket{0000}+ \alpha_2\ket{0001}+\cdots+\alpha_8\ket{1111}$. Because there are only three distinct coefficients present in the UCC factor regardless of the rank, the binary encoding allows us to reduce the size of the ancilla bank logarithmically so that it grows with the rank, not the exponential of the rank. 
\begin{table}[h!]
    \[
    \begin{array}{c}
    \Qcircuit @C=1.2em @R=1.5em{
    \lstick{\ket{0}_1}&\gate{R_{X_1}} &\ctrl{1}&\ctrlo{1}&\ctrlo{1} &\ctrlo{1}&\ctrlo{1}&\ctrlo{1} &\qw \\ 
    \lstick{\ket{0}_2}&\qw &\multigate{2}{H}&\gate{R_{Y_2}}&\ctrl{1}&\ctrlo{1}&\ctrlo{1}&\ctrlo{1} &\qw\\ 
    \lstick{\ket{0}_3}&\qw &\ghost{H}&\qw&\multigate{1}{H}&\gate{R_{Y_3}}&\ctrl{1}&\ctrlo{1} &\qw\\
    \lstick{\ket{0}_4}&\qw &\ghost{H}&\qw&\ghost{H}&\qw&\gate{H} &\gate{R_{Y_4}}&\qw\\}
    \end{array}
    \]
\captionsetup{justification=raggedright}
\caption{Quantum circuit for preparing the ancilla bank of the LCU query for a rank-$2$ UCC factor (so-called doubles).}
\label{table:preparedouble}
\end{table}
A \textsc{Prepare} circuit for the doubles factor is shown in the Tab.~\ref{table:preparedouble}. The $H$ gates are Hadamard operators, and the $R_{X_i}$ and $R_{Y_i}$ gates are rotations by an angle $\Theta_i$ along the $X$ and $Y$ axis, respectively. For a UCC doubles operator, $n=2$, and there are three distinct coefficients: eight terms with $\frac{i\sin{\theta}}{8}$, seven terms with $\frac{\cos{\theta}-1}{8}$, and one term with $\frac{\cos{\theta}+7}{8}$. The four angles used in the circuit shown in Tab.~\ref{table:preparedouble} can be found analytically:
\begin{align}
    &\Theta_1 = \arcsin{\bigg(-\frac{\sqrt{2}}{4}\sin{\theta}\bigg)} \\
    &\Theta_2 = \arcsin{\bigg(\frac{\cos{\theta}-1}{\sqrt{14+2\cos^2{\theta}}}\bigg)}\\
    &\Theta_3 = \arcsin{\bigg(\frac{\sqrt{2}}{2}\frac{\cos{\theta}-1}{\sqrt{13+\cos^2{\theta}+2\cos{\theta}}}\bigg)}\\
    &\Theta_4 = \arcsin{\bigg(\frac{\sqrt{2}}{2}\frac{\cos{\theta}-1}{\sqrt{25+\cos^2{\theta}+6\cos{\theta}}}\bigg)}
\end{align}
And the magnitudes of these four angles are shown in Fig.~ \ref{fig:uccd_prepare_angles}.
The \textsc{Prepare} subroutine is implemented by encoding the amplitudes $1+\frac{\cos\theta - 1}{2^{2n-1}}$ and $\frac{\cos\theta - 1}{2^{2n-1}}$ in $2^{2n-1}$ states that always have one different binary digit than those encoding the amplitude $\frac{i\sin\theta}{2^{2n-1}}$. For example, in the doubles circuit, we encode the amplitudes $1+\frac{\cos\theta - 1}{2^3}$ and $\frac{\cos\theta - 1}{2^3}$ in states $\ket{0000}$ and $\ket{0001} \cdots \ket{0111}$, and encode the amplitude $\frac{i\sin\theta}{2^3}$ in states $\ket{1000} \cdots \ket{1111}$. 

The \textsc{Prepare} circuit for a doubles UCC factor can be straightforwardly generalized to one for a rank-$n$ operator.  In this case, we require $2n$ ancilla qubits, where there are $2^{2n-1}$ subterms with coefficients $\frac{i\sin{\theta}}{2^{2n-1}}$, $2^{2n-1}-1$ subterms with coefficients $\frac{\cos{\theta}-1}{{2^{2n-1}}}$, and one term with coefficient $1+\frac{\cos{\theta}-1}{2^{2n-1}}$. The hierarchical structure of the circuit is a simple generalization of the doubles circuit to higher rank, where the first element, which sets the $\sin$ terms is the same, while the remaining factors are created by extending the hierarchy with multiple controlled Hadamards followed by multiply controlled rotations. The angles for each rotation in the algorithm are
\begin{align}
    &\Theta_1 = \arcsin{\bigg(-\frac{1}{\sqrt{2^{2n-1}} }\sin\theta\bigg)} \\
    &\Theta_k = \arcsin{\bigg(\frac{\cos\theta - 1}{\sqrt{2^{2n-2+k}-2^k+2+2\cos^2\theta +(2^k-4)\cos\theta}} \bigg)}
\end{align} for $2\leq k\leq 2n$.

Most of the quantum computing costs on near-term quantum computers come from the \textsc{Cnot} circuit elements; thus, we focus on counting the number of these gates to estimate the total cost of the circuits. The controlled-$H$ gate, \textsc{CH}, is implemented by one \textsc{Cnot} sandwiched by one $R_Y(\pi/4)$ gate and one $R_Y(-\pi/4)$ gate. The controlled-$R_Y(\theta)$ operator is employed by two \textsc{Cnot} gates and two half rotations. We use the linear-depth method proposed in Ref.~\cite{silva_park_2022} to decompose the multi-qubit control operators into standard \textsc{Cnot} and one-qubit gates. The \textsc{Cnot} cost of an $n$-qubit controlled operator is $8n-12$ for all $n\geq 2$. The circuit needs $2(2n-k+1)(8k-12)$ \textsc{Cnot} gates for each $k$-qubit controlled operation. For $k=1$, the \textsc{Cnot} cost is $2n$. To implement the circuit for an arbitrary rank-$n$, we need to employ the modular sub-circuit shown in Tab.~\ref{table:module} on $2n-1$ ancilla qubits. The upper bound for the number of control qubits used in any module is $2n-1$. The total cost of running the \textsc{Prepare} subroutine is then $2n+2\sum_{k=2} ^{2n-1} (8k-12)(2n+1-k)$, which can be further simplified to $\frac{8}{3}(8n^3-6n^2-\frac{41n}{4}+9)$. Hence, the cost scales like the cube of the rank.
\begin{table}[h!]
    \[
    \begin{array}{c}
    \Qcircuit @C=1.5em @R=1.5em {
    \lstick{\ket{q}_1}& \multigate{5}{\mathcal{M}(\ket{q}_n)}&\qw & & &\ctrlo{2}&\ctrlo{2}&\qw \\
    \vdots& & & & & & & \vdots \\
    \lstick{\ket{q}_m}&\ghost{\mathcal{M}(\ket{q}_n)} & \qw & & &\ctrl{1}&\ctrlo{1}&\qw \\
    \lstick{\ket{q}_n}&\ghost{\mathcal{M}(\ket{q}_n)} &\qw & =  & &\multigate{2}{H}&\gate{R_Y(\Theta)}&\qw \\
    \vdots& & & & & & & \vdots \\
    \lstick{\ket{q}_l}&\ghost{\mathcal{M}(\ket{q}_n)} &\qw &  & &\ghost{H}&\qw&\qw
    }
    \end{array}
    \]
\captionsetup{justification=raggedright}
\caption{Quantum circuit for the modules of qubit-$\ket{q}_n$ used in the \textsc{Prepare} subroutine. The Hadamard operator is anti-controlled by the qubits between $\ket{q}_1$ and $\ket{q}_m$ and controlled by the qubit $\ket{q}_m$. The $R_Y(\Theta)$ operator is anti-controlled by qubits $\ket{q}_1$ to $\ket{q}_m$.}
\label{table:module}
\end{table}

\begin{table}[h!]
    \[
    \begin{array}{c}
    \Qcircuit @C=1em @R=1.5em{
    \lstick{\ket{0}_1}&\gate{R_{X_1}} &\multigate{4}{\mathcal{M}(\ket{0}_2)}&\multigate{4}{\mathcal{M}(\ket{0}_3)}&\qw&\cdots& &\multigate{4}{\mathcal{M}(\ket{0}_l)}&\qw\\ 
    \lstick{\ket{0}_2}&\qw &\ghost{\mathcal{M}(\ket{0}_2)}&\ghost{\mathcal{M}(\ket{0}_3)}&\qw&\cdots& &\ghost{\mathcal{M}(\ket{0}_l)}&\qw\\ 
    \lstick{\ket{0}_3}&\qw &\ghost{\mathcal{M}(\ket{0}_2)}&\ghost{\mathcal{M}(\ket{0}_3)}&\qw&\cdots& &\ghost{\mathcal{M}(\ket{0}_l)}&\qw\\
    \vdots \\
    \lstick{\ket{0}_l}&\qw &\ghost{\mathcal{M}(\ket{0}_2)}&\ghost{\mathcal{M}(\ket{0}_3)}&\qw&\cdots& &\ghost{\mathcal{M}(\ket{0}_l)}&\qw
    }
    \end{array}
    \]
\captionsetup{justification=raggedright}
\caption{Quantum circuit for the \textsc{Prepare} subroutine of arbitrary rank with modules introduced in table \ref{table:module}.}
\label{table:generalwithmodules}
\end{table}

\subsection{\textsc{Select}(\texorpdfstring{$\hat U$}{Lg}) subroutine for rank-2 factors}\label{sec:select}
In this section, we introduce a quantum circuit for implementing the $\textsc{Select}(\hat U)$ operation, illustrated in Eq.~(\ref{eq:selectdef}), for the UCCD factors, where the unitary $\hat U$ is the operator given in Eq.~(\ref{eq:su2jw}). The first step of the \textsc{Select}($\hat U$) circuit is to create one of the Pauli strings from the pool of the excitation operators ($XY$ strings), and one of the Pauli strings from the projection pool ($IZ$ strings). In this example, we opt to create the $Y_l  X_k  X_j  X_i$ and $I_l Z_k Z_j I_i$ strings as shown in Tab.~\ref{table:selectstep1}, however, any arbitrary Pauli string can be the candidate for this step. It is important to note that we use control operations for the $Y_l  X_k  X_j  X_i$ terms, whereas we use anticontrol operations for the $I_l Z_k Z_j I_i$ terms. This is because we are partitioning the ancilla bank into two sectors, one part for the $XY$ strings, and one part for the $IZ$ strings. In this case, ancilla-bank states $\ket{1000}$ to $\ket{1111}$ are used to create the $XY$ strings, and ancilla-bank states $\ket{0000}$ to $\ket{0111}$ are used to create the $IZ$ strings. The resulting state, omitting the corresponding coefficients, which are prepared in the previous step, becomes
\begin{align}
    &(\ket{0000} + \cdots + \ket{0111})\ket{Z_k Z_j}\nonumber \\
    + &(\ket{1000}+\cdots + \ket{1111})\bigotimes _{a=l+1} ^{k-1} Z_{a}\bigotimes _{b=j+1} ^{i-1} Z_{b}\ket{Y_l  X_k  X_j  X_i}
\end{align}
\begin{table}[h!]
    \[
    \begin{array}{c}
    \Qcircuit @C=1.2em @R=1.2em{
    \lstick{\ket{0}_1} & \ctrl{1}&\ctrl{5}&\ctrl{1}  &\ctrl{3} &\ctrl{5}& \ctrl{7} &\ctrlo{3} &\ctrlo{5}&\qw \\ 
    \lstick{l} &\multigate{2}{Z_{k-1} ^{l+1}}&\qw&\gate{Y}  &\qw&\qw&\qw&\qw&\qw&\qw \\ 
    \vdots \\
    \lstick{k} &\ghost{Z_{k-1} ^{l+1}} &\qw&\qw  &\gate{X}&\qw &\qw&\gate{Z}&\qw&\qw\\
    \vdots \\
    \lstick{j} &\qw &\multigate{2}{Z_{i-1} ^{j+1}}&\qw&\qw  &\gate{X}&\qw &\qw&\gate{Z}&\qw\\
    \vdots \\
    \lstick{i} &\qw  &\ghost{Z_{i-1} ^{j+1}}&\qw&\qw &\qw & \gate{X}&\qw&\qw&\qw\\
    }
    \end{array}
    \]
\captionsetup{justification=raggedright}
\caption{Circuit to create $Y_l  X_k  X_j  X_i$ and $I_l Z_k Z_j I_i$. $\ket{0}_1$ denotes the first qubit of the ancilla bank. The first two circuit components are the controlled Pauli $Z$ gates applied on qubits between (exclusively) indices $l$ and $k$, and between indices $j$ and $i$.}
\label{table:selectstep1}
\end{table}

The construction of the controlled Pauli $Z$ gates shown in Tab.~(\ref{table:selectstep1}) is described next.
\begin{table}[h!]
    \[
    \begin{array}{c}
    \Qcircuit @C=1.5em @R=1.5em {
    & \ctrl{1}&\qw & & &\ctrl{2}&\ctrl{3}&\qw \\
    \lstick{l} &\multigate{3}{Z_{i-1} ^{j+1}} & \qw & & &\qw&\qw&\qw \\
    &\ghost{Z_{i-1} ^{j+1}} &\qw & =  & &\gate{Z}&\qw&\qw \\
    &\ghost{Z_{i-1} ^{j+1}} &\qw &   & &\qw&\gate{Z}&\qw\\
    \lstick{k} &\ghost{Z_{i-1} ^{j+1}} & \qw& & &\qw&\qw&\qw
    }
    \end{array}
    \]
\captionsetup{justification=raggedright}
\caption{The circuit for implementing the boxed controlled Pauli $Z$ operators shown in table (\ref{table:selectstep1}). The Pauli $Z$ operators are being applied to the qubits $l+1$ to $k-1$.}
\end{table}
With the starting reference states prepared, we can then create the entire state exactly. The first step is to apply a single-qubit controlled Pauli $Z$ operator on qubits $l$ and $k$, where the control is to be conditioned on the last qubit of the ancilla bank. The second step is similar, in that a single-qubit controlled Pauli $Z$ operator is applied on qubits $j$ and $i$. The control qubit of the second step is the second to last qubit of the ancilla bank. The final step is to apply the single-qubit controlled Pauli $Z$ on qubits $k$ and $j$, with the control qubit  being the second qubit of the ancilla bank. The circuit diagram is illustrated in Tab.~\ref{table:selectstep2}.
\begin{table}[h!]
    \[
    \begin{array}{c}
    \Qcircuit @C=1.6em @R=1.2em{
    \lstick{\ket{0}_2}&\qw&\qw&\qw&\qw&\ctrl{5}&\ctrl{7}&\qw\\
    \lstick{\ket{0}_3}&\qw&\qw&\ctrl{6}&\ctrl{8}&\qw&\qw&\qw\\
    \lstick{\ket{0}_4} & \ctrl{1}&\ctrl{3}&\qw  &\qw &\qw& \qw &\qw \\ 
    \lstick{l} &\gate{Z}&\qw&\qw  &\qw&\qw&\qw&\qw \\ 
    \vdots \\
    \lstick{k} &\qw&\gate{Z}&\qw  &\qw &\gate{Z} &\qw&\qw\\
    \vdots \\
    \lstick{j} &\qw &\qw&\gate{Z}&\qw  &\qw&\gate{Z} &\qw\\
    \vdots \\
    \lstick{i} &\qw  &\qw&\qw&\gate{Z} &\qw & \qw&\qw\\
    }
    \end{array}
    \]
\captionsetup{justification=raggedright}
\caption{Circuit to create the state shown in equation (\ref{eq:su2jw}). $\ket{0}_i$ denotes the $i$th qubit of the ancilla bank.}
\label{table:selectstep2}
\end{table}
The Pauli strings and their corresponding states in the ancilla bank are shown in Tab.~\ref{table:correspondingtable}.
\begin{table}[h!]
    \centering
    \begin{tabular}{c|c|c}
        $\ket{0 0 0}_{2,3,4}$ & $\ket{0}_1 = 1$ & $\ket{0}_1 = 0$\\
        \hline
        $\ket{000}$ &$YXXX$ & $IZZI$\\
        $\ket{001}$ &$XYXX$ & $ZIZI$\\
        $\ket{010}$ &$YXYY$ & $IZIZ$\\
        $\ket{011}$ &$XYYY$ & $ZIIZ$\\
        $\ket{100}$ &$YYYX$ & $IIII$\\
        $\ket{101}$ &$XXYX$ & $ZZII$\\
        $\ket{110}$ &$YYXY$ & $IIZZ$\\
        $\ket{111}$ &$XXXY$ & $ZZZZ$\\
    \end{tabular}
    \caption{The $16$ Pauli strings created by schemes shown in Tabs.~\ref{table:selectstep1} and \ref{table:selectstep2}, and their associated binary encodings in the ancilla bank. }
    \label{table:correspondingtable}
\end{table}
The qubits on which the control operations are conditioned are chosen specifically for this table. In practice, when applying this algorithm, one needs to predetermine a table similar to Tab.~\ref{table:correspondingtable} for all the binary encodings and their corresponding Pauli substrings, and choose accordingly the starting reference states and the control qubits to be used in the approach illustrated shown in Tabs.~\ref{table:selectstep1} and \ref{table:selectstep2}.
\subsection{\textsc{Select}(\texorpdfstring{$\hat U$}{Lg}) for arbitrary rank-n}
In this section, we demonstrate the algorithm for the rank-$n$ UCC factor buy generalizing the algorithms shown in Secs.~\ref{sec:prepare} and \ref{sec:select}. First, let us re-examine the case for rank-$2$; that is, the doubles. Define groups $\mathbf{G_1} = \{G_{11},G_{12}\}$ and $\mathbf{G_2} = \{G_{21},G_{22}\}$, with elements $G = \sigma_{i} \otimes \sigma_{j}$, where $\sigma_{i}$ and $\sigma_j$ are two different Pauli operators acting on different qubits. Additionally, the group elements have the following identities
\begin{align}
    &G_{11} \cdot (\sigma_{z}\otimes \sigma_{z}) = G_{12}, G_{12} \cdot (\sigma_{z}\otimes \sigma_{z}) = G_{11} \label{eq:group1}\\
    &G_{11} \cdot (\mathds{1}\otimes \sigma_{z}) = G_{22}, G_{11} \cdot (\sigma_{z}\otimes \mathds{1}) = G_{21} \label{eq:group2}\\
    &G_{21} \cdot (\sigma_{z}\otimes \sigma_{z}) = G_{22}, G_{22} \cdot (\sigma_{z}\otimes \sigma_{z}) = G_{21} \label{eq:group3}\\
    &G_{21} \cdot(\mathds{1}\otimes \sigma_{z}) = G_{12}, G_{21} \cdot (\sigma_{z}\otimes \mathds{1}) = G_{11}\label{eq:group4},
\end{align}
where $\sigma_{z}$ is the Pauli $Z$ operator being applied on qubit $a_1$. It should be clear that the expression in Eq.~(\ref{eq:su2jw}) is of the schematic form $\sum_{p\neq q}\sum_r G_{pr}\otimes G_{qr} + G^\prime_{rp}\otimes G^\prime_{rq}$, omitting the coefficients, where the $\mathbf{G}$ groups contain $XY$ subterms and the $\mathbf{G^\prime}$ contain $IZ$ subterms. In the example shown in the Secs.~\ref{sec:prepare} and \ref{sec:select}, the corresponding groups are $\mathbf{G_1} = \{G_{11}=YX, G_{12}=XY \}$, $\mathbf{G_2} = \{G_{21}=XX, G_{22}=YY \}$, $\mathbf{G_1^\prime} = \{G_{11}^\prime=IZ, G_{12}^\prime=ZI \}$, and $\mathbf{G_2^\prime} = \{G_{21}^\prime=II, G_{22}^\prime=ZZ \}$. There are in total three steps needed to create the eigenfunction, hence we opted to use three digits for the binary encoding in the ancilla bank.
\begin{table}[h!]
    \centering
    \begin{tabular}{c|cccc}
        $\otimes$&$G_{11}$ & $G_{12}$ &$G_{21}$ &$G_{22}$ \\
        \hline
        $G_{11}$ & 0&0 &\cellcolor[HTML]{89CFF0}$G_{11}G_{21}$&\cellcolor[HTML]{F4C2C2}$G_{11}G_{22}$\\
        $G_{12}$ &0 &0 &\cellcolor[HTML]{FFE135}$G_{12}G_{21}$&\cellcolor[HTML]{F4C2C2}$G_{12}G_{22}$\\
        $G_{21}$ &\cellcolor[HTML]{7FFFD0}$G_{21}G_{11}$&\cellcolor[HTML]{7FFFD0}$G_{21}G_{12}$&0 &0 \\
        $G_{22}$ &\cellcolor[HTML]{7FFFD0}$G_{22}G_{11}$&\cellcolor[HTML]{7FFFD0}$G_{22}G_{12}$&0 &0 \\
    \end{tabular}
    \caption{Matrix that shows the scheme for the $XY$ subterms presented in the Sec.~\ref{sec:select}. The entry colored in blue, $G_{11}G_{21}$ is the initial reference state. After the first step, the yellow entry $G_{12}G_{21}$ is created. The pink entries are created after the second step. The first two steps use the identities in Eqs.~(\ref{eq:group1}) and (\ref{eq:group3}). The final step is to use the identities in Eqs.~(\ref{eq:group2}) and (\ref{eq:group4}) to create the green block.}
    \label{table:doublestable}
\end{table}

\begin{table}[h!]
    \centering
    \begin{tabular}{c|cccc}
        $\otimes$&$G_{11}^\prime$ & $G_{12}^\prime$ &$G_{21}^\prime$ &$G_{22}^\prime$ \\
        \hline
        $G_{11}^\prime$ & \cellcolor[HTML]{F4C2C2}$G_{11}^\prime G_{11}^\prime$&\cellcolor[HTML]{89CFF0}$G_{11}^\prime G_{12}^\prime$&0&0\\
        $G_{12}^\prime$ &\cellcolor[HTML]{F4C2C2}$G_{12}^\prime G_{11}^\prime$&\cellcolor[HTML]{FFE135}$G_{12}^\prime G_{12}^\prime$&0&0\\
        $G_{21}^\prime$ &0&0&\cellcolor[HTML]{7FFFD0}$G_{21}^\prime G_{21}^\prime$&\cellcolor[HTML]{7FFFD0}$G_{21}^\prime G_{22}^\prime$ \\
        $G_{22}^\prime$ &0&0&\cellcolor[HTML]{7FFFD0}$G_{22}^\prime G_{21}^\prime$ &\cellcolor[HTML]{7FFFD0}$G_{22}^\prime G_{22}^\prime$ \\
    \end{tabular}
    \caption{Matrix that shows the scheme for the $IZ$ subterms presented in Sec.~\ref{sec:select}. The entry colored in blue, $G_{11}^\prime G_{12}^\prime$ is chosen to be the starting reference state. After the first step, the yellow entry $G_{12}^\prime G_{12}^\prime$ is created. The pink entries result after the second step. The first two steps use the identities in Eqs.~(\ref{eq:group1}) and (\ref{eq:group3}). The final step is to use the identities in Eqs.~(\ref{eq:group2}) and (\ref{eq:group4}) to create the entire green block.}
    \label{table:doublestableprime}
\end{table}

Similar to the way we define for the rank-$2$ factors, for rank-$3$, we define groups $\mathbf{G_1} = \{G_{11},G_{12},G_{13},G_{14}\}$ and $\mathbf{G_2} = \{G_{21},G_{22},G_{23},G_{24}\}$, therefore the wavefunction for the UCC triples factors after the JW transformation take the form $\sum_{p\neq q}\sum_r G_{pr}\otimes G_{qr} + G^\prime_{rp}\otimes G^\prime_{rq}$. The set of identities for rank-$3$ is
\begin{align}
    &G_{11}  (\sigma_{z}\otimes \sigma_{z}\otimes \mathds{1}) = G_{14}, G_{14}  (\sigma_{z}\otimes \sigma_{z}\otimes \mathds{1}) = G_{11} \label{eq:triplegroup1}\\
    &G_{12}  (\sigma_{z}\otimes \sigma_{z}\otimes \mathds{1}) = G_{13}, G_{13}  (\sigma_{z}\otimes \sigma_{z}\otimes \mathds{1}) = G_{12} \label{eq:triplegroup9}\\
    &G_{11}  (\mathds{1}\otimes \sigma_{z}\otimes \sigma_{z}) = G_{12}, G_{12}  (\mathds{1}\otimes \sigma_{z}\otimes \sigma_{z}) = G_{11} \label{eq:triplegroup2}\\
    &G_{13}  (\mathds{1}\otimes \sigma_{z}\otimes \sigma_{z}) = G_{14}, G_{14}  (\mathds{1}\otimes \sigma_{z}\otimes \sigma_{z}) = G_{13} \label{eq:triplegroup10}\\
    &G_{21}  (\sigma_{z}\otimes \sigma_{z}\otimes \mathds{1}) = G_{24}, G_{24}  (\sigma_{z}\otimes \sigma_{z}\otimes \mathds{1}) = G_{21} \label{eq:triplegroup3}\\
    &G_{22}  (\sigma_{z}\otimes \sigma_{z}\otimes \mathds{1}) = G_{23}, G_{23}  (\sigma_{z}\otimes \sigma_{z}\otimes \mathds{1}) = G_{24} \label{eq:triplegroup11}\\
    &G_{21}  (\mathds{1}\otimes \sigma_{z}\otimes \sigma_{z}) = G_{22}, G_{22}  (\mathds{1}\otimes \sigma_{z}\otimes \sigma_{z}) = G_{21} \label{eq:triplegroup4}\\
    &G_{23}  (\mathds{1}\otimes \sigma_{z}\otimes \sigma_{z}) = G_{24}, G_{24}  (\mathds{1}\otimes \sigma_{z}\otimes \sigma_{z}) = G_{22} \label{eq:triplegroup12}\\
    &G_{11}  (\mathds{1}\otimes \mathds{1}\otimes \sigma_{z}) = G_{21}, G_{12}  (\mathds{1}\otimes \mathds{1}\otimes \sigma_{z}) = G_{22} \label{eq:triplegroup5}\\
    &G_{13}  (\mathds{1}\otimes \mathds{1}\otimes \sigma_{z}) = G_{23}, G_{14}  (\mathds{1}\otimes \mathds{1}\otimes \sigma_{z}) = G_{24} \label{eq:triplegroup6}\\
    &G_{21}  (\sigma_{z}\otimes\mathds{1}\otimes \mathds{1}) = G_{13}, G_{22}  (\sigma_{z}\otimes\mathds{1}\otimes \mathds{1}) = G_{14} \label{eq:triplegroup7}\\
    &G_{23}  (\sigma_{z}\otimes\mathds{1}\otimes \mathds{1}) = G_{11}, G_{24}  (\sigma_{z}\otimes\mathds{1}\otimes \mathds{1}) = G_{12} \label{eq:triplegroup8}.
\end{align}
Shown in the Tab.~\ref{table:triplestable}, there are in total five steps needed to create the exact JW-transformed unitary, thus we use five out of six digits for the binary encoding in the ancilla bank. 
\begin{table*}[h!]
    \centering
    \begin{tabular}{c|cccccccc}
         $\otimes$&$G_{11}$ & $G_{12}$ &$G_{13}$ &$G_{14}$&$G_{21}$ & $G_{22}$ &$G_{23}$ &$G_{24}$ \\
        \hline
        $G_{11}$ & 0&0 &0&0&\cellcolor[HTML]{89CFF0}$G_{11}G_{21}$&\cellcolor[HTML]{ED872D}$G_{11}G_{22}$&\cellcolor[HTML]{ED872D}$G_{11}G_{23}$&\cellcolor[HTML]{F4C2C2}$G_{11}G_{24}$\\
        $G_{12}$ &0 &0& 0&0&\cellcolor[HTML]{FFE135}$G_{12}G_{21}$&\cellcolor[HTML]{ED872D}$G_{12}G_{22}$&\cellcolor[HTML]{ED872D}$G_{12}G_{23}$&\cellcolor[HTML]{F4C2C2}$G_{12}G_{24}$\\
        $G_{13}$ &0 &0& 0&0&\cellcolor[HTML]{FFE135}$G_{13}G_{21}$&\cellcolor[HTML]{ED872D}$G_{13}G_{22}$&\cellcolor[HTML]{ED872D}$G_{13}G_{23}$&\cellcolor[HTML]{F4C2C2}$G_{13}G_{24}$\\
        $G_{14}$ &0 &0& 0&0&\cellcolor[HTML]{91A3B0}$G_{14}G_{21}$&\cellcolor[HTML]{ED872D}$G_{14}G_{22}$&\cellcolor[HTML]{ED872D}$G_{14}G_{23}$&\cellcolor[HTML]{F4C2C2}$G_{14}G_{24}$\\
        $G_{21}$ &\cellcolor[HTML]{7FFFD0}$G_{21}G_{11}$&\cellcolor[HTML]{7FFFD0}$G_{21}G_{12}$&\cellcolor[HTML]{7FFFD0}$G_{21}G_{13}$&\cellcolor[HTML]{7FFFD0}$G_{21}G_{14}$&0 &0& 0&0 \\
        $G_{22}$ &\cellcolor[HTML]{7FFFD0}$G_{22}G_{11}$&\cellcolor[HTML]{7FFFD0}$G_{22}G_{12}$&\cellcolor[HTML]{7FFFD0}$G_{22}G_{13}$&\cellcolor[HTML]{7FFFD0}$G_{22}G_{14}$&0 &0& 0&0 \\
        $G_{23}$ &\cellcolor[HTML]{7FFFD0}$G_{23}G_{11}$&\cellcolor[HTML]{7FFFD0}$G_{23}G_{12}$&\cellcolor[HTML]{7FFFD0}$G_{23}G_{13}$&\cellcolor[HTML]{7FFFD0}$G_{23}G_{14}$&0 &0& 0&0 \\
        $G_{24}$ &\cellcolor[HTML]{7FFFD0}$G_{24}G_{11}$&\cellcolor[HTML]{7FFFD0}$G_{24}G_{12}$&\cellcolor[HTML]{7FFFD0}$G_{24}G_{13}$&\cellcolor[HTML]{7FFFD0}$G_{24}G_{14}$&0 &0& 0&0 
    \end{tabular}
    \caption{Matrix illustrating the scheme for constructing $XY$ substrings in the JW-transformed unitary of the UCCT factors. The blue entry is chosen to be the starting reference state. After the first step using identity (\ref{eq:triplegroup1}), grey entry $G_{14}G_{21}$ is created. Using the identities in Eqs.~(\ref{eq:triplegroup9}) and (\ref{eq:triplegroup2}), the yellow entries are created after the second step. The orange and pink columns are created after the next two steps via the identities in Eqs.~(\ref{eq:triplegroup10}) through (\ref{eq:triplegroup4}). The final step is to apply the identities in Eqs.~(\ref{eq:triplegroup5}) through (\ref{eq:triplegroup8}) to create the entire green block.}
    \label{table:triplestable}
\end{table*}

\begin{table*}[h!]
    \centering
    \begin{tabular}{c|cccccccc}
         $\otimes$&$G_{11}^\prime $ & $G_{12}^\prime $ &$G_{13}^\prime $ &$G_{14}^\prime $&$G_{21}^\prime $ & $G_{22}^\prime $ &$G_{23}^\prime $ &$G_{24}^\prime $ \\
        \hline
        $G_{11}^\prime $ & \cellcolor[HTML]{F4C2C2}$G_{11}^\prime G_{11}^\prime $&\cellcolor[HTML]{ED872D}$G_{11}^\prime G_{12}^\prime $&\cellcolor[HTML]{ED872D}$G_{11}^\prime G_{13}^\prime $&\cellcolor[HTML]{89CFF0}$G_{11}^\prime G_{14}^\prime $&0&0&0&0\\
        $G_{12}^\prime $ &\cellcolor[HTML]{F4C2C2}$G_{12}^\prime G_{11}^\prime $&\cellcolor[HTML]{ED872D}$G_{12}^\prime G_{12}^\prime $& \cellcolor[HTML]{ED872D}$G_{12}^\prime G_{13}^\prime $&\cellcolor[HTML]{FFE135}$G_{12}^\prime G_{14}^\prime $&0&0&0&0\\
        $G_{13}^\prime $ &\cellcolor[HTML]{F4C2C2}$G_{13}^\prime G_{11}^\prime $&\cellcolor[HTML]{ED872D}$G_{13}^\prime G_{23}^\prime $& \cellcolor[HTML]{ED872D}$G_{13}^\prime G_{13}^\prime $&\cellcolor[HTML]{FFE135}$G_{13}^\prime G_{14}^\prime $&0&0&0&0\\
        $G_{14}^\prime $ &\cellcolor[HTML]{F4C2C2}$G_{14}^\prime G_{11}^\prime $&\cellcolor[HTML]{ED872D}$G_{14}^\prime G_{12}^\prime $&\cellcolor[HTML]{ED872D}$G_{14}^\prime G_{13}^\prime $&\cellcolor[HTML]{91A3B0}$G_{14}^\prime G_{14}^\prime $&0&0&0&0\\
        $G_{21}^\prime $ &0 &0& 0&0&\cellcolor[HTML]{7FFFD0}$G_{21}^\prime G_{21}^\prime $&\cellcolor[HTML]{7FFFD0}$G_{21}^\prime G_{22}^\prime $&\cellcolor[HTML]{7FFFD0}$G_{21}^\prime G_{23}^\prime $&\cellcolor[HTML]{7FFFD0}$G_{21}^\prime G_{24}^\prime $\\
        $G_{22}^\prime $ &0 &0& 0&0&\cellcolor[HTML]{7FFFD0}$G_{22}^\prime G_{21}^\prime $&\cellcolor[HTML]{7FFFD0}$G_{22}^\prime G_{22}^\prime $&\cellcolor[HTML]{7FFFD0}$G_{22}^\prime G_{23}^\prime $&\cellcolor[HTML]{7FFFD0}$G_{22}^\prime G_{24}^\prime $\\
        $G_{23}^\prime $ &0 &0& 0&0&\cellcolor[HTML]{7FFFD0}$G_{23}^\prime G_{21}^\prime $&\cellcolor[HTML]{7FFFD0}$G_{23}^\prime G_{22}^\prime $&\cellcolor[HTML]{7FFFD0}$G_{23}^\prime G_{23}^\prime $&\cellcolor[HTML]{7FFFD0}$G_{23}^\prime G_{24}^\prime $\\
        $G_{24}^\prime $ &0 &0& 0&0&\cellcolor[HTML]{7FFFD0}$G_{24}^\prime G_{21}^\prime $&\cellcolor[HTML]{7FFFD0}$G_{24}^\prime G_{22}^\prime $&\cellcolor[HTML]{7FFFD0}$G_{24}^\prime G_{23}^\prime $&\cellcolor[HTML]{7FFFD0}$G_{24}^\prime G_{24}^\prime $\\
    \end{tabular}
    \caption{Matrix illustrating the scheme for constructing $IZ$ substrings in the JW-transformed unitary of the UCCT factors. The blue entry is the starting reference state. After the first step using the identity in Eq.~(\ref{eq:triplegroup1}), the grey entry $G_{14}G_{14}$ is created. Using the identities in Eqs.~(\ref{eq:triplegroup9}) and (\ref{eq:triplegroup2}), the yellow entries are created after the second step. The orange and pink columns are created after the next two steps via the identities in Eqs.~(\ref{eq:triplegroup10}) through (\ref{eq:triplegroup4}). The final step is to apply the identities in Eqs.~(\ref{eq:triplegroup5}) through (\ref{eq:triplegroup8}) to create the entire green block.}
    \label{table:triplestableprime}
\end{table*}

For a UCC factor with arbitrary rank $n$, a total number of $n-1$ transformations are required to complete the first column of the matrix. An additional $n-1$ transformations are then required to complete the rest of the diagonal or off-diagonal block, depending on whether the subterms are $XY$ strings or $IZ$ strings. The final step is to perform the 'flip' transformation to make the entire matrix. Therefore, for any UCC factor of rank $n$, a total number of $2n-1$ steps are needed to construct the JW-transformed unitary operator.

\begin{table*}[h!]
    \centering
    \begin{tabular}{c|cccccccc}
         $\otimes$&$G_{21}$ & $G_{22}$ &$G_{23}$ &$G_{24}$&$G_{25}$ & $G_{26}$ &$G_{27}$ &$G_{28}$ \\
        \hline
        $G_{11}$ &\cellcolor[HTML]{89CFF0}$G_{11}G_{21}$&\cellcolor[HTML]{ED872D}$G_{11}G_{22}$&\cellcolor[HTML]{F4BBFF}$G_{11}G_{23}$&\cellcolor[HTML]{F4BBFF}$G_{11}G_{24}$&\cellcolor[HTML]{F4BBFF}$G_{11}G_{25}$&\cellcolor[HTML]{F4BBFF}$G_{11}G_{26}$&\cellcolor[HTML]{ED872D}$G_{11}G_{27}$&\cellcolor[HTML]{F4C2C2}$G_{11}G_{28}$\\
        $G_{12}$ &\cellcolor[HTML]{7FFFD0}$G_{12}G_{21}$&\cellcolor[HTML]{ED872D}$G_{12}G_{22}$&\cellcolor[HTML]{F4BBFF}$G_{12}G_{23}$&\cellcolor[HTML]{F4BBFF}$G_{12}G_{24}$&\cellcolor[HTML]{F4BBFF}$G_{12}G_{25}$&\cellcolor[HTML]{F4BBFF}$G_{12}G_{26}$&\cellcolor[HTML]{ED872D}$G_{12}G_{27}$&\cellcolor[HTML]{F4C2C2}$G_{12}G_{28}$\\
        $G_{13}$ &\cellcolor[HTML]{FFE135}$G_{13}G_{21}$&\cellcolor[HTML]{ED872D}$G_{13}G_{22}$&\cellcolor[HTML]{F4BBFF}$G_{13}G_{23}$&\cellcolor[HTML]{F4BBFF}$G_{13}G_{24}$&\cellcolor[HTML]{F4BBFF}$G_{13}G_{25}$&\cellcolor[HTML]{F4BBFF}$G_{13}G_{26}$&\cellcolor[HTML]{ED872D}$G_{13}G_{27}$&\cellcolor[HTML]{F4C2C2}$G_{13}G_{28}$\\
        $G_{14}$ &\cellcolor[HTML]{FFE135}$G_{14}G_{21}$&\cellcolor[HTML]{ED872D}$G_{14}G_{22}$&\cellcolor[HTML]{F4BBFF}$G_{14}G_{23}$&\cellcolor[HTML]{F4BBFF}$G_{14}G_{24}$&\cellcolor[HTML]{F4BBFF}$G_{14}G_{25}$&\cellcolor[HTML]{F4BBFF}$G_{14}G_{26}$&\cellcolor[HTML]{ED872D}$G_{14}G_{27}$&\cellcolor[HTML]{F4C2C2}$G_{14}G_{28}$\\
        $G_{15}$ &\cellcolor[HTML]{FFE135}$G_{15}G_{21}$&\cellcolor[HTML]{ED872D}$G_{15}G_{22}$&\cellcolor[HTML]{F4BBFF}$G_{15}G_{23}$&\cellcolor[HTML]{F4BBFF}$G_{15}G_{24}$&\cellcolor[HTML]{F4BBFF}$G_{15}G_{25}$&\cellcolor[HTML]{F4BBFF}$G_{15}G_{26}$&\cellcolor[HTML]{ED872D}$G_{15}G_{27}$&\cellcolor[HTML]{F4C2C2}$G_{15}G_{28}$\\
        $G_{16}$ &\cellcolor[HTML]{FFE135}$G_{16}G_{21}$&\cellcolor[HTML]{ED872D}$G_{16}G_{22}$&\cellcolor[HTML]{F4BBFF}$G_{16}G_{23}$&\cellcolor[HTML]{F4BBFF}$G_{16}G_{24}$&\cellcolor[HTML]{F4BBFF}$G_{16}G_{25}$&\cellcolor[HTML]{F4BBFF}$G_{16}G_{26}$&\cellcolor[HTML]{ED872D}$G_{16}G_{27}$&\cellcolor[HTML]{F4C2C2}$G_{16}G_{28}$\\
        $G_{17}$ &\cellcolor[HTML]{7FFFD0}$G_{17}G_{21}$&\cellcolor[HTML]{ED872D}$G_{17}G_{22}$&\cellcolor[HTML]{F4BBFF}$G_{17}G_{23}$&\cellcolor[HTML]{F4BBFF}$G_{17}G_{24}$&\cellcolor[HTML]{F4BBFF}$G_{17}G_{25}$&\cellcolor[HTML]{F4BBFF}$G_{17}G_{26}$&\cellcolor[HTML]{ED872D}$G_{17}G_{27}$&\cellcolor[HTML]{F4C2C2}$G_{17}G_{28}$\\
        $G_{18}$ &\cellcolor[HTML]{B2BEB5}$G_{18}G_{21}$&\cellcolor[HTML]{ED872D}$G_{18}G_{22}$&\cellcolor[HTML]{F4BBFF}$G_{18}G_{23}$&\cellcolor[HTML]{F4BBFF}$G_{18}G_{24}$&\cellcolor[HTML]{F4BBFF}$G_{18}G_{25}$&\cellcolor[HTML]{F4BBFF}$G_{18}G_{26}$&\cellcolor[HTML]{ED872D}$G_{18}G_{27}$&\cellcolor[HTML]{F4C2C2}$G_{18}G_{28}$\\
    \end{tabular}
    \caption{Off-diagonal block of a matrix illustrating the scheme for constructing $XY$ substrings in the JW-transformed unitary of the UCCQ factors. The blue entry of the first column is chosen to be the starting reference state. The grey entry is created after the first step, followed by the green entries in the second step and the yellow entries in the third step. The pink entries of the last column are created after the fourth step, followed by the orange columns in the fifth step and purple columns in the sixth step.}
    \label{table:quadoffblock}
\end{table*}

\subsection{Gate counts}
The LCU framework for rank-$n$ UCC factors is given by a simple circuit implementation of the JW-transformed unitary in the form of 
\begin{equation}
    \sum_{p\neq q}\sum_r G_{pr}\otimes G_{qr} + G^\prime_{rp}\otimes G^\prime_{rq},
\end{equation}
where each group $\mathbf{G_{p}}$ contains $2^{n-1}$ elements that commute with every other element in the same group but anti-commute with elements in the other group. The total number of steps in the \textsc{Select}($\hat U$) subroutine is $2n-1$, for rank-$n$  factors. Within each step of the \textsc{Select}($\hat U$) subroutine, two single-qubit controlled Pauli $Z$ operators are needed, making the total number of \textsc{Cnot} gates $4n-2$. In addition, $4n+\sum_{i}^{2n-2} \rho_i$ \textsc{Cnot} gates are needed for initializing the reference Jordan-Wigner strings. Here,  $\rho_i$ is the number of qubits between the $i$th pair of the active orbitals.  In the case of a UCCD factor discussed in Secs.~\ref{sec:prepare} and \ref{sec:select}, $\rho_1$ is the number of qubits between qubits $k$ and $l$ and $\rho_2$ is the number of qubits between qubits $j$ and $i$. The circuit for preparing the ancilla bank hosts the majority of the complexity, where the total \textsc{Cnot} cost is $\frac{8}{3}(8n^3-6n^2-\frac{41n}{4}+9)$. The total number of \textsc{Cnot} gates used in the LCU circuit for preparing the JW-transformed unitary is then $6\cdot (\frac{8}{3}(8n^3-6n^2-\frac{41n}{4}+9))+3\cdot (8n-2+\sum_{i}^{2n-2}\rho_i)$, which can be further simplified to $128n^3-96n^2-140n+138+3\sum_{i}^{2n-2}\rho_i$. Note that this has a large prefactor in the scaling with the rank. The total number of ancilla qubits required for this framework is $2n$. For a UCC factor with an arbitrary set of active orbitals, the \textsc{Cnot} cost of the circuits proposed in our work eventually becomes favorable compared with other existing methods when the rank becomes large ($n\geq 9$), including the one proposed in the authors' previous work \cite{xu_lee_freericks_2022}, although for low-rank factors, the fermionic-excitation-based (FEB) algorithm proposed by Ref. \cite{magoulas_evangelista_2023} is more efficient.

\begin{figure}
    \centering
    \includegraphics[width=0.45\textwidth]{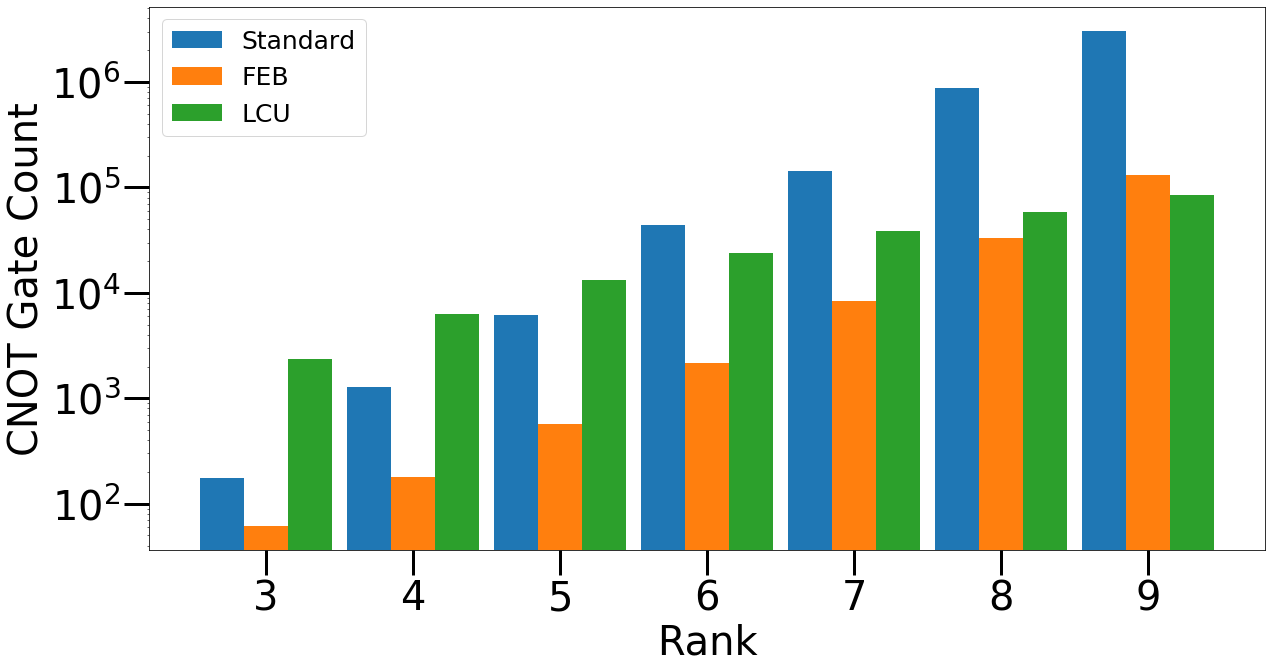}
    \caption{\textsc{Cnot} gate counts for three different algorithms, the standard \textsc{Cnot} cascading circuits, the FEB circuits \cite{magoulas_evangelista_2023}, and the linear-combination-of-unitary query circuits introduced in this work.}
    \label{fig:cnotcomparison}
\end{figure}

\section{Discussion}
In summary, we have introduced an application of the LCU query model that efficiently simulates the factorized UCC ansatze with a scaling that goes like the cube of the rank. We have demonstrated the quantum circuits for the \textsc{Prepare} and the \textsc{Select}($\hat U$) subroutines, whose \textsc{Cnot} counts scale linearly with the rank of the UCC ansatze and the number of active spin-orbitals. The \textsc{Prepare} operator employs a quantum circuit that scales as the cube of the rank of the UCC factor and requires a linear number of ancilla. The proposed LCU framework greatly reduces the number of two-qubit gates for high-rank UCC factors, which are needed for simulating strongly correlated systems on NISQ devices and future fault-tolerant quantum computers \cite{freericks_2022}. Alternatively, a low-rank factorization method does exist for the electronic Hamiltonian and the UCC operator \cite{motta_ye_mcclean_li_minnich_babbush_chan_2021}, but it is not clear how well it extends to high rank, so we do not compare the scaling of their method with ours. Circuits that implement the \textsc{Select}($\hat U$) subroutine for more general Jordan-Wigner strings with linear scaling have also been developed \cite{wan_2021}. Another approach that optimizes the \textsc{Cnot} cascading circuits in Ref. (\cite{romero_babbush_mcclean_hempel_love_aspuru-guzik_2018}) for low-rank factors have been recently developed as well \cite{wang_li_monroe_nam_2021, cowtan_dilkes_duncan_simmons_sivarajah_2020, magoulas_evangelista_2023}.

\section{Acknowledgements}
We acknowledge helpful discussions with Joseph T. Lee. L. Xu and J. K. Freericks were supported by the U.S. Department of Energy, Office of Science, Office of Advanced Scientific Computing Research (ASCR), Quantum Computing Application Teams (QCATS) program, under field work proposal number ERKJ347. J. K. Freericks was also supported by the McDevitt bequest at Georgetown University.
\bibliography{biblio}

\begin{thebibliography}{29}%
\makeatletter
\providecommand \@ifxundefined [1]{%
 \@ifx{#1\undefined}
}%
\providecommand \@ifnum [1]{%
 \ifnum #1\expandafter \@firstoftwo
 \else \expandafter \@secondoftwo
 \fi
}%
\providecommand \@ifx [1]{%
 \ifx #1\expandafter \@firstoftwo
 \else \expandafter \@secondoftwo
 \fi
}%
\providecommand \natexlab [1]{#1}%
\providecommand \enquote  [1]{``#1''}%
\providecommand \bibnamefont  [1]{#1}%
\providecommand \bibfnamefont [1]{#1}%
\providecommand \citenamefont [1]{#1}%
\providecommand \href@noop [0]{\@secondoftwo}%
\providecommand \href [0]{\begingroup \@sanitize@url \@href}%
\providecommand \@href[1]{\@@startlink{#1}\@@href}%
\providecommand \@@href[1]{\endgroup#1\@@endlink}%
\providecommand \@sanitize@url [0]{\catcode `\\12\catcode `\$12\catcode
  `\&12\catcode `\#12\catcode `\^12\catcode `\_12\catcode `\%12\relax}%
\providecommand \@@startlink[1]{}%
\providecommand \@@endlink[0]{}%
\providecommand \url  [0]{\begingroup\@sanitize@url \@url }%
\providecommand \@url [1]{\endgroup\@href {#1}{\urlprefix }}%
\providecommand \urlprefix  [0]{URL }%
\providecommand \Eprint [0]{\href }%
\providecommand \doibase [0]{https://doi.org/}%
\providecommand \selectlanguage [0]{\@gobble}%
\providecommand \bibinfo  [0]{\@secondoftwo}%
\providecommand \bibfield  [0]{\@secondoftwo}%
\providecommand \translation [1]{[#1]}%
\providecommand \BibitemOpen [0]{}%
\providecommand \bibitemStop [0]{}%
\providecommand \bibitemNoStop [0]{.\EOS\space}%
\providecommand \EOS [0]{\spacefactor3000\relax}%
\providecommand \BibitemShut  [1]{\csname bibitem#1\endcsname}%
\let\auto@bib@innerbib\@empty
\bibitem [{\citenamefont {Aspuru-Guzik}(2005)}]{aspuru-guzik_2005}%
  \BibitemOpen
  \bibfield  {author} {\bibinfo {author} {\bibfnamefont {A.}~\bibnamefont
  {Aspuru-Guzik}},\ }\href {https://doi.org/10.1126/science.1113479} {\bibfield
   {journal} {\bibinfo  {journal} {Science}\ }\textbf {\bibinfo {volume}
  {309}},\ \bibinfo {pages} {1704–1707} (\bibinfo {year} {2005})}\BibitemShut
  {NoStop}%
\bibitem [{\citenamefont {Whitfield}\ \emph {et~al.}(2011)\citenamefont
  {Whitfield}, \citenamefont {Biamonte},\ and\ \citenamefont
  {Aspuru-Guzik}}]{whitfield_biamonte_aspuru-guzik_2011}%
  \BibitemOpen
  \bibfield  {author} {\bibinfo {author} {\bibfnamefont {J.~D.}\ \bibnamefont
  {Whitfield}}, \bibinfo {author} {\bibfnamefont {J.}~\bibnamefont
  {Biamonte}},\ and\ \bibinfo {author} {\bibfnamefont {A.}~\bibnamefont
  {Aspuru-Guzik}},\ }\href {https://doi.org/10.1080/00268976.2011.552441}
  {\bibfield  {journal} {\bibinfo  {journal} {Mol. Phys.}\ }\textbf {\bibinfo
  {volume} {109}},\ \bibinfo {pages} {735–750} (\bibinfo {year}
  {2011})}\BibitemShut {NoStop}%
\bibitem [{\citenamefont {Sherrill}\ and\ \citenamefont
  {Schaefer}(1999)}]{fci}%
  \BibitemOpen
  \bibfield  {author} {\bibinfo {author} {\bibfnamefont {C.~D.}\ \bibnamefont
  {Sherrill}}\ and\ \bibinfo {author} {\bibfnamefont {H.~F.}\ \bibnamefont
  {Schaefer}},\ }\href {https://doi.org/10.1016/s0065-3276(08)60532-8}
  {\bibfield  {journal} {\bibinfo  {journal} {Adv. Quantum Chem.}\ }\textbf
  {\bibinfo {volume} {34}},\ \bibinfo {pages} {143–269} (\bibinfo {year}
  {1999})}\BibitemShut {NoStop}%
\bibitem [{\citenamefont {Bartlett}\ and\ \citenamefont
  {Purvis}(1978)}]{bartlett_purvis_1978}%
  \BibitemOpen
  \bibfield  {author} {\bibinfo {author} {\bibfnamefont {R.~J.}\ \bibnamefont
  {Bartlett}}\ and\ \bibinfo {author} {\bibfnamefont {G.~D.}\ \bibnamefont
  {Purvis}},\ }\href {https://doi.org/10.1002/qua.560140504} {\bibfield
  {journal} {\bibinfo  {journal} {Int. J. Quantum Chem.}\ }\textbf {\bibinfo
  {volume} {14}},\ \bibinfo {pages} {561–581} (\bibinfo {year}
  {1978})}\BibitemShut {NoStop}%
\bibitem [{\citenamefont {Bartlett}\ and\ \citenamefont {Musiał}(2007)}]{cc}%
  \BibitemOpen
  \bibfield  {author} {\bibinfo {author} {\bibfnamefont {R.~J.}\ \bibnamefont
  {Bartlett}}\ and\ \bibinfo {author} {\bibfnamefont {M.}~\bibnamefont
  {Musiał}},\ }\href {https://doi.org/10.1103/revmodphys.79.291} {\bibfield
  {journal} {\bibinfo  {journal} {Rev. Mod. Phys.}\ }\textbf {\bibinfo {volume}
  {79}},\ \bibinfo {pages} {291–352} (\bibinfo {year} {2007})}\BibitemShut
  {NoStop}%
\bibitem [{\citenamefont {Purvis}\ and\ \citenamefont
  {Bartlett}(1982)}]{purvis_bartlett_1982}%
  \BibitemOpen
  \bibfield  {author} {\bibinfo {author} {\bibfnamefont {G.~D.}\ \bibnamefont
  {Purvis}}\ and\ \bibinfo {author} {\bibfnamefont {R.~J.}\ \bibnamefont
  {Bartlett}},\ }\href {https://doi.org/10.1063/1.443164} {\bibfield  {journal}
  {\bibinfo  {journal} {J. Chem. Phys.}\ }\textbf {\bibinfo {volume} {76}},\
  \bibinfo {pages} {1910–1918} (\bibinfo {year} {1982})}\BibitemShut
  {NoStop}%
\bibitem [{\citenamefont {Anand}\ \emph {et~al.}(2022)\citenamefont {Anand},
  \citenamefont {Schleich}, \citenamefont {Alperin-Lea}, \citenamefont
  {Jensen}, \citenamefont {Sim}, \citenamefont {Díaz-Tinoco}, \citenamefont
  {Kottmann}, \citenamefont {Degroote}, \citenamefont {Izmaylov}, \citenamefont
  {Aspuru-Guzik},\ and\ \citenamefont {et~al.}}]{aspuruguzik_uccreview}%
  \BibitemOpen
  \bibfield  {author} {\bibinfo {author} {\bibfnamefont {A.}~\bibnamefont
  {Anand}}, \bibinfo {author} {\bibfnamefont {P.}~\bibnamefont {Schleich}},
  \bibinfo {author} {\bibfnamefont {S.}~\bibnamefont {Alperin-Lea}}, \bibinfo
  {author} {\bibfnamefont {P.~W.}\ \bibnamefont {Jensen}}, \bibinfo {author}
  {\bibfnamefont {S.}~\bibnamefont {Sim}}, \bibinfo {author} {\bibfnamefont
  {M.}~\bibnamefont {Díaz-Tinoco}}, \bibinfo {author} {\bibfnamefont {J.~S.}\
  \bibnamefont {Kottmann}}, \bibinfo {author} {\bibfnamefont {M.}~\bibnamefont
  {Degroote}}, \bibinfo {author} {\bibfnamefont {A.~F.}\ \bibnamefont
  {Izmaylov}}, \bibinfo {author} {\bibfnamefont {A.}~\bibnamefont
  {Aspuru-Guzik}},\ and\ \bibinfo {author} {\bibnamefont {et~al.}},\ }\href
  {https://doi.org/10.1039/d1cs00932j} {\bibfield  {journal} {\bibinfo
  {journal} {Chem. Soc. Rev.}\ }\textbf {\bibinfo {volume} {51}},\ \bibinfo
  {pages} {1659–1684} (\bibinfo {year} {2022})}\BibitemShut {NoStop}%
\bibitem [{\citenamefont {Peruzzo}\ \emph {et~al.}(2014)\citenamefont
  {Peruzzo}, \citenamefont {Mcclean}, \citenamefont {Shadbolt}, \citenamefont
  {Yung}, \citenamefont {Zhou}, \citenamefont {Love}, \citenamefont
  {Aspuru-Guzik},\ and\ \citenamefont {O’Brien}}]{vqe}%
  \BibitemOpen
  \bibfield  {author} {\bibinfo {author} {\bibfnamefont {A.}~\bibnamefont
  {Peruzzo}}, \bibinfo {author} {\bibfnamefont {J.}~\bibnamefont {Mcclean}},
  \bibinfo {author} {\bibfnamefont {P.}~\bibnamefont {Shadbolt}}, \bibinfo
  {author} {\bibfnamefont {M.-H.}\ \bibnamefont {Yung}}, \bibinfo {author}
  {\bibfnamefont {X.-Q.}\ \bibnamefont {Zhou}}, \bibinfo {author}
  {\bibfnamefont {P.~J.}\ \bibnamefont {Love}}, \bibinfo {author}
  {\bibfnamefont {A.}~\bibnamefont {Aspuru-Guzik}},\ and\ \bibinfo {author}
  {\bibfnamefont {J.~L.}\ \bibnamefont {O’Brien}},\ }\href
  {https://doi.org/10.1038/ncomms5213} {\bibfield  {journal} {\bibinfo
  {journal} {Nat. Commun.}\ }\textbf {\bibinfo {volume} {5}},\ \bibinfo {pages}
  {4213} (\bibinfo {year} {2014})}\BibitemShut {NoStop}%
\bibitem [{\citenamefont {Preskill}(2018)}]{preskill_2018}%
  \BibitemOpen
  \bibfield  {author} {\bibinfo {author} {\bibfnamefont {J.}~\bibnamefont
  {Preskill}},\ }\href {https://doi.org/10.22331/q-2018-08-06-79} {\bibfield
  {journal} {\bibinfo  {journal} {Quantum}\ }\textbf {\bibinfo {volume} {2}},\
  \bibinfo {pages} {79} (\bibinfo {year} {2018})}\BibitemShut {NoStop}%
\bibitem [{\citenamefont {Szabo}\ and\ \citenamefont
  {Ostlund}(2006)}]{szabo_ostlund_2006}%
  \BibitemOpen
  \bibfield  {author} {\bibinfo {author} {\bibfnamefont {A.}~\bibnamefont
  {Szabo}}\ and\ \bibinfo {author} {\bibfnamefont {N.~S.}\ \bibnamefont
  {Ostlund}},\ }\href@noop {} {\emph {\bibinfo {title} {Modern quantum
  chemistry: introduction to advanced electronic structure theory}}}\ (\bibinfo
   {publisher} {Dover Publications, Mineola, NY},\ \bibinfo {year}
  {2006})\BibitemShut {NoStop}%
\bibitem [{\citenamefont {Taketa}\ \emph {et~al.}(1966)\citenamefont {Taketa},
  \citenamefont {Huzinaga},\ and\ \citenamefont
  {O-Ohata}}]{taketa_huzinaga_o-ohata_1966}%
  \BibitemOpen
  \bibfield  {author} {\bibinfo {author} {\bibfnamefont {H.}~\bibnamefont
  {Taketa}}, \bibinfo {author} {\bibfnamefont {S.}~\bibnamefont {Huzinaga}},\
  and\ \bibinfo {author} {\bibfnamefont {K.}~\bibnamefont {O-Ohata}},\ }\href
  {https://doi.org/10.1143/jpsj.21.2313} {\bibfield  {journal} {\bibinfo
  {journal} {J. Phys. Soc. Japan}\ }\textbf {\bibinfo {volume} {21}},\ \bibinfo
  {pages} {2313–2324} (\bibinfo {year} {1966})}\BibitemShut {NoStop}%
\bibitem [{\citenamefont {Shavitt}\ and\ \citenamefont
  {Bartlett}(2009)}]{shavitt_bartlett_2009}%
  \BibitemOpen
  \bibfield  {author} {\bibinfo {author} {\bibfnamefont {I.}~\bibnamefont
  {Shavitt}}\ and\ \bibinfo {author} {\bibfnamefont {R.~J.}\ \bibnamefont
  {Bartlett}},\ }\href@noop {} {\emph {\bibinfo {title} {Many-body methods in
  chemistry and physics: MBPT and coupled-cluster theory}}}\ (\bibinfo
  {publisher} {Cambridge University Press},\ \bibinfo {year}
  {2009})\BibitemShut {NoStop}%
\bibitem [{\citenamefont {Bartlett}\ \emph {et~al.}(1989)\citenamefont
  {Bartlett}, \citenamefont {Kucharski},\ and\ \citenamefont
  {Noga}}]{bartlett_kucharski_noga_1989}%
  \BibitemOpen
  \bibfield  {author} {\bibinfo {author} {\bibfnamefont {R.~J.}\ \bibnamefont
  {Bartlett}}, \bibinfo {author} {\bibfnamefont {S.~A.}\ \bibnamefont
  {Kucharski}},\ and\ \bibinfo {author} {\bibfnamefont {J.}~\bibnamefont
  {Noga}},\ }\href {https://doi.org/10.1016/s0009-2614(89)87372-5} {\bibfield
  {journal} {\bibinfo  {journal} {Chem. Phys. Lett.}\ }\textbf {\bibinfo
  {volume} {155}},\ \bibinfo {pages} {133–140} (\bibinfo {year}
  {1989})}\BibitemShut {NoStop}%
\bibitem [{\citenamefont {Schaefer}(2013)}]{schaefer_2013}%
  \BibitemOpen
  \bibfield  {author} {\bibinfo {author} {\bibfnamefont {H.~F.}\ \bibnamefont
  {Schaefer}},\ }\href@noop {} {\emph {\bibinfo {title} {Methods of electronic
  structure theory}}}\ (\bibinfo  {publisher} {Springer Science Business Media,
  LLC, New York, NY},\ \bibinfo {year} {2013})\BibitemShut {NoStop}%
\bibitem [{\citenamefont {Cooper}\ and\ \citenamefont
  {Knowles}(2010)}]{cooper_knowles_2010}%
  \BibitemOpen
  \bibfield  {author} {\bibinfo {author} {\bibfnamefont {B.}~\bibnamefont
  {Cooper}}\ and\ \bibinfo {author} {\bibfnamefont {P.~J.}\ \bibnamefont
  {Knowles}},\ }\href {https://doi.org/10.1063/1.3520564} {\bibfield  {journal}
  {\bibinfo  {journal} {J. Chem. Phys.}\ }\textbf {\bibinfo {volume} {133}},\
  \bibinfo {pages} {234102} (\bibinfo {year} {2010})}\BibitemShut {NoStop}%
\bibitem [{\citenamefont {Chen}\ \emph {et~al.}(2021)\citenamefont {Chen},
  \citenamefont {Cheng},\ and\ \citenamefont
  {Freericks}}]{chen_cheng_freericks_2021}%
  \BibitemOpen
  \bibfield  {author} {\bibinfo {author} {\bibfnamefont {J.}~\bibnamefont
  {Chen}}, \bibinfo {author} {\bibfnamefont {H.-P.}\ \bibnamefont {Cheng}},\
  and\ \bibinfo {author} {\bibfnamefont {J.~K.}\ \bibnamefont {Freericks}},\
  }\href {https://doi.org/10.1021/acs.jctc.0c01052} {\bibfield  {journal}
  {\bibinfo  {journal} {J. Chem. Theory Comput.}\ }\textbf {\bibinfo {volume}
  {17}},\ \bibinfo {pages} {841–847} (\bibinfo {year} {2021})}\BibitemShut
  {NoStop}%
\bibitem [{\citenamefont {Romero}\ \emph {et~al.}(2018)\citenamefont {Romero},
  \citenamefont {Babbush}, \citenamefont {Mcclean}, \citenamefont {Hempel},
  \citenamefont {Love},\ and\ \citenamefont
  {Aspuru-Guzik}}]{romero_babbush_mcclean_hempel_love_aspuru-guzik_2018}%
  \BibitemOpen
  \bibfield  {author} {\bibinfo {author} {\bibfnamefont {J.}~\bibnamefont
  {Romero}}, \bibinfo {author} {\bibfnamefont {R.}~\bibnamefont {Babbush}},
  \bibinfo {author} {\bibfnamefont {J.~R.}\ \bibnamefont {Mcclean}}, \bibinfo
  {author} {\bibfnamefont {C.}~\bibnamefont {Hempel}}, \bibinfo {author}
  {\bibfnamefont {P.~J.}\ \bibnamefont {Love}},\ and\ \bibinfo {author}
  {\bibfnamefont {A.}~\bibnamefont {Aspuru-Guzik}},\ }\href
  {https://doi.org/10.1088/2058-9565/aad3e4} {\bibfield  {journal} {\bibinfo
  {journal} {Quantum Sci. and Technol.}\ }\textbf {\bibinfo {volume} {4}},\
  \bibinfo {pages} {014008} (\bibinfo {year} {2018})}\BibitemShut {NoStop}%
\bibitem [{\citenamefont {Barkoutsos}\ \emph {et~al.}(2018)\citenamefont
  {Barkoutsos}, \citenamefont {Gonthier}, \citenamefont {Sokolov},
  \citenamefont {Moll}, \citenamefont {Salis}, \citenamefont {Fuhrer},
  \citenamefont {Ganzhorn}, \citenamefont {Egger}, \citenamefont {Troyer},
  \citenamefont {Mezzacapo},\ and\ \citenamefont {et~al.}}]{barkoutsos_2018}%
  \BibitemOpen
  \bibfield  {author} {\bibinfo {author} {\bibfnamefont {P.~K.}\ \bibnamefont
  {Barkoutsos}}, \bibinfo {author} {\bibfnamefont {J.~F.}\ \bibnamefont
  {Gonthier}}, \bibinfo {author} {\bibfnamefont {I.}~\bibnamefont {Sokolov}},
  \bibinfo {author} {\bibfnamefont {N.}~\bibnamefont {Moll}}, \bibinfo {author}
  {\bibfnamefont {G.}~\bibnamefont {Salis}}, \bibinfo {author} {\bibfnamefont
  {A.}~\bibnamefont {Fuhrer}}, \bibinfo {author} {\bibfnamefont
  {M.}~\bibnamefont {Ganzhorn}}, \bibinfo {author} {\bibfnamefont {D.~J.}\
  \bibnamefont {Egger}}, \bibinfo {author} {\bibfnamefont {M.}~\bibnamefont
  {Troyer}}, \bibinfo {author} {\bibfnamefont {A.}~\bibnamefont {Mezzacapo}},\
  and\ \bibinfo {author} {\bibnamefont {et~al.}},\ }\bibfield  {journal}
  {\bibinfo  {journal} {Phys. Rev. A}\ }\textbf {\bibinfo {volume} {98}},\
  \href {https://doi.org/10.1103/physreva.98.022322}
  {10.1103/physreva.98.022322} (\bibinfo {year} {2018})\BibitemShut {NoStop}%
\bibitem [{\citenamefont {Xu}\ \emph {et~al.}(2022)\citenamefont {Xu},
  \citenamefont {Lee},\ and\ \citenamefont
  {Freericks}}]{xu_lee_freericks_2022}%
  \BibitemOpen
  \bibfield  {author} {\bibinfo {author} {\bibfnamefont {L.}~\bibnamefont
  {Xu}}, \bibinfo {author} {\bibfnamefont {J.~T.}\ \bibnamefont {Lee}},\ and\
  \bibinfo {author} {\bibfnamefont {J.~K.}\ \bibnamefont {Freericks}},\
  }\bibfield  {journal} {\bibinfo  {journal} {Phys. Rev. A}\ }\textbf {\bibinfo
  {volume} {105}},\ \href {https://doi.org/10.1103/physreva.105.012406}
  {10.1103/physreva.105.012406} (\bibinfo {year} {2022})\BibitemShut {NoStop}%
\bibitem [{\citenamefont {Xu}\ \emph {et~al.}(2020)\citenamefont {Xu},
  \citenamefont {Lee},\ and\ \citenamefont
  {Freericks}}]{xu_lee_freericks_2020}%
  \BibitemOpen
  \bibfield  {author} {\bibinfo {author} {\bibfnamefont {L.}~\bibnamefont
  {Xu}}, \bibinfo {author} {\bibfnamefont {J.~T.}\ \bibnamefont {Lee}},\ and\
  \bibinfo {author} {\bibfnamefont {J.~K.}\ \bibnamefont {Freericks}},\ }\href
  {https://doi.org/10.1142/s0217984920400497} {\bibfield  {journal} {\bibinfo
  {journal} {Mod. Phys. Lett. B}\ }\textbf {\bibinfo {volume} {34}},\ \bibinfo
  {pages} {2040049} (\bibinfo {year} {2020})}\BibitemShut {NoStop}%
\bibitem [{\citenamefont {Freericks}(2022)}]{freericks_2022}%
  \BibitemOpen
  \bibfield  {author} {\bibinfo {author} {\bibfnamefont {J.~K.}\ \bibnamefont
  {Freericks}},\ }\href {https://doi.org/10.3390/sym14030494} {\bibfield
  {journal} {\bibinfo  {journal} {Symmetry}\ }\textbf {\bibinfo {volume}
  {14}},\ \bibinfo {pages} {494} (\bibinfo {year} {2022})}\BibitemShut
  {NoStop}%
\bibitem [{\citenamefont {Childs}\ and\ \citenamefont
  {Wiebe}(2012)}]{childs_wiebe_2012}%
  \BibitemOpen
  \bibfield  {author} {\bibinfo {author} {\bibfnamefont {A.~M.}\ \bibnamefont
  {Childs}}\ and\ \bibinfo {author} {\bibfnamefont {N.}~\bibnamefont {Wiebe}},\
  }\href {https://doi.org/10.26421/qic12.11-12-1} {\bibfield  {journal}
  {\bibinfo  {journal} {Quantum Inf. Comput.}\ }\textbf {\bibinfo {volume}
  {12}},\ \bibinfo {pages} {901–924} (\bibinfo {year} {2012})}\BibitemShut
  {NoStop}%
\bibitem [{\citenamefont {Berry}\ \emph {et~al.}(2015)\citenamefont {Berry},
  \citenamefont {Childs}, \citenamefont {Cleve}, \citenamefont {Kothari},\ and\
  \citenamefont {Somma}}]{berry_childs_cleve_kothari_somma_2015}%
  \BibitemOpen
  \bibfield  {author} {\bibinfo {author} {\bibfnamefont {D.~W.}\ \bibnamefont
  {Berry}}, \bibinfo {author} {\bibfnamefont {A.~M.}\ \bibnamefont {Childs}},
  \bibinfo {author} {\bibfnamefont {R.}~\bibnamefont {Cleve}}, \bibinfo
  {author} {\bibfnamefont {R.}~\bibnamefont {Kothari}},\ and\ \bibinfo {author}
  {\bibfnamefont {R.~D.}\ \bibnamefont {Somma}},\ }\bibfield  {journal}
  {\bibinfo  {journal} {Phys. Rev. Lett.}\ }\textbf {\bibinfo {volume} {114}},\
  \href {https://doi.org/10.1103/physrevlett.114.090502}
  {10.1103/physrevlett.114.090502} (\bibinfo {year} {2015})\BibitemShut
  {NoStop}%
\bibitem [{\citenamefont {Wan}(2021)}]{wan_2021}%
  \BibitemOpen
  \bibfield  {author} {\bibinfo {author} {\bibfnamefont {K.}~\bibnamefont
  {Wan}},\ }\href {https://doi.org/10.22331/q-2021-01-12-380} {\bibfield
  {journal} {\bibinfo  {journal} {Quantum}\ }\textbf {\bibinfo {volume} {5}},\
  \bibinfo {pages} {380} (\bibinfo {year} {2021})}\BibitemShut {NoStop}%
\bibitem [{\citenamefont {da~Silva}\ and\ \citenamefont
  {Park}(2022)}]{silva_park_2022}%
  \BibitemOpen
  \bibfield  {author} {\bibinfo {author} {\bibfnamefont {A.~J.}\ \bibnamefont
  {da~Silva}}\ and\ \bibinfo {author} {\bibfnamefont {D.~K.}\ \bibnamefont
  {Park}},\ }\bibfield  {journal} {\bibinfo  {journal} {Phys. Rev. A}\ }\textbf
  {\bibinfo {volume} {106}},\ \href
  {https://doi.org/10.1103/physreva.106.042602} {10.1103/physreva.106.042602}
  (\bibinfo {year} {2022})\BibitemShut {NoStop}%
\bibitem [{\citenamefont {Magoulas}\ and\ \citenamefont
  {Evangelista}(2023)}]{magoulas_evangelista_2023}%
  \BibitemOpen
  \bibfield  {author} {\bibinfo {author} {\bibfnamefont {I.}~\bibnamefont
  {Magoulas}}\ and\ \bibinfo {author} {\bibfnamefont {F.~A.}\ \bibnamefont
  {Evangelista}},\ }\bibfield  {journal} {\bibinfo  {journal} {J. Chem. Theory
  Comput.}\ }\href {https://doi.org/10.1021/acs.jctc.2c01016}
  {10.1021/acs.jctc.2c01016} (\bibinfo {year} {2023})\BibitemShut {NoStop}%
\bibitem [{\citenamefont {Motta}\ \emph {et~al.}(2021)\citenamefont {Motta},
  \citenamefont {Ye}, \citenamefont {McClean}, \citenamefont {Li},
  \citenamefont {Minnich}, \citenamefont {Babbush},\ and\ \citenamefont
  {Chan}}]{motta_ye_mcclean_li_minnich_babbush_chan_2021}%
  \BibitemOpen
  \bibfield  {author} {\bibinfo {author} {\bibfnamefont {M.}~\bibnamefont
  {Motta}}, \bibinfo {author} {\bibfnamefont {E.}~\bibnamefont {Ye}}, \bibinfo
  {author} {\bibfnamefont {J.~R.}\ \bibnamefont {McClean}}, \bibinfo {author}
  {\bibfnamefont {Z.}~\bibnamefont {Li}}, \bibinfo {author} {\bibfnamefont
  {A.~J.}\ \bibnamefont {Minnich}}, \bibinfo {author} {\bibfnamefont
  {R.}~\bibnamefont {Babbush}},\ and\ \bibinfo {author} {\bibfnamefont
  {G.~K.-L.}\ \bibnamefont {Chan}},\ }\bibfield  {journal} {\bibinfo  {journal}
  {npj Quantum Inf.}\ }\textbf {\bibinfo {volume} {7}},\ \href
  {https://doi.org/10.1038/s41534-021-00416-z} {10.1038/s41534-021-00416-z}
  (\bibinfo {year} {2021})\BibitemShut {NoStop}%
\bibitem [{\citenamefont {Wang}\ \emph {et~al.}(2021)\citenamefont {Wang},
  \citenamefont {Li}, \citenamefont {Monroe},\ and\ \citenamefont
  {Nam}}]{wang_li_monroe_nam_2021}%
  \BibitemOpen
  \bibfield  {author} {\bibinfo {author} {\bibfnamefont {Q.}~\bibnamefont
  {Wang}}, \bibinfo {author} {\bibfnamefont {M.}~\bibnamefont {Li}}, \bibinfo
  {author} {\bibfnamefont {C.}~\bibnamefont {Monroe}},\ and\ \bibinfo {author}
  {\bibfnamefont {Y.}~\bibnamefont {Nam}},\ }\href
  {https://doi.org/10.22331/q-2021-07-26-509} {\bibfield  {journal} {\bibinfo
  {journal} {Quantum}\ }\textbf {\bibinfo {volume} {5}},\ \bibinfo {pages}
  {509} (\bibinfo {year} {2021})}\BibitemShut {NoStop}%
\bibitem [{\citenamefont {Cowtan}\ \emph {et~al.}(2020)\citenamefont {Cowtan},
  \citenamefont {Dilkes}, \citenamefont {Duncan}, \citenamefont {Simmons},\
  and\ \citenamefont
  {Sivarajah}}]{cowtan_dilkes_duncan_simmons_sivarajah_2020}%
  \BibitemOpen
  \bibfield  {author} {\bibinfo {author} {\bibfnamefont {A.}~\bibnamefont
  {Cowtan}}, \bibinfo {author} {\bibfnamefont {S.}~\bibnamefont {Dilkes}},
  \bibinfo {author} {\bibfnamefont {R.}~\bibnamefont {Duncan}}, \bibinfo
  {author} {\bibfnamefont {W.}~\bibnamefont {Simmons}},\ and\ \bibinfo {author}
  {\bibfnamefont {S.}~\bibnamefont {Sivarajah}},\ }\href
  {https://doi.org/10.4204/eptcs.318.13} {\bibfield  {journal} {\bibinfo
  {journal} {Electron. Proc. Theor. Comput. Sci.}\ }\textbf {\bibinfo {volume}
  {318}},\ \bibinfo {pages} {213–228} (\bibinfo {year} {2020})}\BibitemShut
  {NoStop}%
\end{thebibliography}%
\bibliographystyle{apsrev4-2}

\end{document}